\documentclass{aa}  

\usepackage{xcolor}
\usepackage{graphicx}
\usepackage{txfonts}
\usepackage{amsmath}
\usepackage{yhmath}
\usepackage{bm}
\def\f{{Fig.~}}
\def\fs{{Figs.~}}
\def\ff{{Figure~}}
\def\s{{Sect.~}}

\def\e{{Eq.~}}
\def\es{{Eqs.~}}
\def\t{{Table~}}

\def\Gyr{{\rm Gyr}}
\def\kpc{{\rm kpc}}
\def\kms{{\rm km\,s^{-1}}}
\def\kmskpc{{\rm \,km\,s^{-1}\,kpc^{-1}}}

\def\Gaia{{\it Gaia }}
\def\Gaiaf{{\it Gaia}}
\def\ok2{{$\omega-\frac{1}{2}\kappa$}}
\def\okfour{{$\omega-\frac{1}{4}\kappa$}}

\begin{document}

   \title{Tidally induced spiral arm wraps encoded in phase space}


   \author{T. Antoja
          \inst{1,2,3}
          \and
         P. Ramos\inst{1,2,3,4}
          \and
          F. López-Guitart\inst{1,2}
          \and
           F. Anders\inst{1,2,3}
        \and
      M. Bernet\inst{1,2,3}
          \and
          C. Laporte\inst{1,2,3}
          }
  \institute{Departament de Física Qu\`antica i Astrof\'isica (FQA), Universitat de Barcelona (UB),  c. Mart\'i i Franqu\`es, 1, 08028 Barcelona, Spain
           \email{tantoja@fqa.ub.edu}
    \and{Institut de Ci\`encies del Cosmos (ICCUB), Universitat de Barcelona (UB), c. Mart\'i i Franqu\`es, 1, 08028 Barcelona, Spain}
    \and{Institut d'Estudis Espacials de Catalunya (IEEC), c. Gran Capit\`a, 2-4, 08034 Barcelona, Spain} 
         \and{
             Observatoire astronomique de Strasbourg, Universit{\'e} de Strasbourg, CNRS, 11 rue de l’Universit{\'e}, 67000 Strasbourg, France}\\
             }

   \date{Received September 15, 1996; accepted March 16, 1997}

 
  \abstract
      {The impact of Sagittarius and other satellite galaxies such as the Large Magellanic Cloud on our Galaxy and in particular its disc is gradually being disclosed. Simulations tailored to the interaction of the Milky Way (MW) and Sagittarius show rings and spiral arms appearing in the Galaxy disc. However, spiral arms can also be induced by the bar or by disc instabilities.}
      {We aim to study the dynamics of tidally induced spiral arms in the context of the different encounters with Sagittarius and determine their kinematic signatures in the shape of ridges and waves in angular momentum, similar to those detected with Gaia DR2.}
      {We built toy models of the interaction between a host and a satellite galaxy using orbital integrations after a tidal distant encounter. We derived analytically the shape of the structures seen in phase space as a function of time for simple power-law potential models. We compared these models to a more realistic N-body simulation of the MW Sagittarius-like interaction and also to real data from Gaia DR3.}
      {As previously found, an impulsive distant tidal approach of a galactic satellite generates a kick in velocities that leads to a two-armed spiral structure. The arms are made of orbits in between their apocentres and pericentres, thus, they correspond to regions with average negative galactocentric radial velocity. The two-arm pattern rotates at an angular speed of $\omega-1/2\kappa$ which depends on Galactocentric radius, thus causing a wind-up with time. This winding produces ridges in the $R$-$V_\phi$ projection with alternating signs of $V_R$ and oscillations of $V_R$ in the $L_Z$-$\phi$ space, similar to those observed in the Gaia data. The frequency of these kinematic features increases with time, offering a powerful means to infer the potential and the perturbation's onset time and azimuthal phase. Fourier analysis allows us to date the impact times of simple models and even to date perturbations induced from subsequent pericentric passages that appear as simultaneous waves. For the MW, the Fourier analysis indicates a superposition of two different frequencies, confirming previous studies. Assuming that both are due to impulsive and distant pericentre passages, we find perturbation times <0.6 Gyr and in the range of 0.8-2.1 Gyr. The latter is compatible with a previous pericentre of Sagittarius and would be associated to about four wraps of the spiral arms in the observed radial range.}
      {Further work on the self-gravitating response of galactic discs and possible degeneracies with secular processes induced by the bar is necessary. Our study is a first  step towards shedding more light on the elusive structure and dynamics of the spiral arms of the Galaxy.}


   \maketitle


\section{Introduction}

The hierarchical formation of galaxies in the paradigm of cold dark matter has consequences, not only on how galaxies grow, but also on how they evolve dynamically \citep[e.g.][]{Dressler1980,Barnes1992}. Undeniable evidence of this are the many galaxies that have been caught interacting with a companion \citep{Arp1966}. The Milky Way (MW) is no exception: the impact of Sagittarius and other galactic satellites such as the Large Magellanic Cloud (LMC) on our Galaxy and, in particular, on its disc is gradually unfolding. For example, host-satellite interactions have been proposed as a mechanism of excitation of the  Galactic warp  \citep[e.g.][]{Hunter1969,Weinberg1995,Ibata1998} and disc heating \citep{Quinn1993}. The vertical asymmetries in the density and in the vertical velocity seen in \citet{Widrow2012} also raised suspicion of disc perturbations from satellites \citep{Gomez2013}. More recent observational evidence for this process came from the discovery of the phase spiral \citep{Antoja2018} in the second data release \citep[DR2;][]{Brown2018} of the \Gaia mission \citep{Prusti2016}. 

Mergers can alter the morphology of galactic discs, exciting the formation of bridge and tail arms in the outermost regions, and of bars, rings, and spiral structure in the inner parts \citep[e.g.][]{Toomre1972,Noguchi1987,Barnes1992,Kazantzidis2008,Pettitt2018}. Simulations tailored to the interaction of the MW and Sagittarius also show that rings and spiral arms would appear in our disc as a response to the interaction \citep{Younger2008,Purcell2011}. On the other hand, spiral arms can also be excited internally by bars or can be self-excited disc modes. 

There are multiple theories about the dynamics, formation, and persistence of spiral structure (e.g. density wave model, swing-amplification theory, transient structures, invariant manifolds; see reviews by \citealt{Sellwood2021a} and \citealt{Dobbs2014}). For the particular case of the MW, there is evidence of spiral arms in young and old populations \citep{Georgelin1976,Reid2009,Drimmel2001}, but a precise global map of the spiral structure is missing, 
let alone a solid theory for its origin and dynamics. Persistent unknown aspects are the number of spiral arms in the Galaxy, their exact composition 
 (gaseous versus stellar), their long- or short-lived nature, their excitation mechanism and relation with the bar, and how they rotate with respect to the stellar disc (as a rigid pattern, with pattern speed depending on radius, or co-rotating with the stars; \citealt{RocaFabrega2013}). 
Despite all these unknowns, there are fewer doubts about the importance of spiral structure in the dynamical evolution of the stellar disc, its gas, and the newly formed stars \citep[e.g.][]{Debattista2006,Sellwood2014}. 

Density structures such as spiral arms and rings, independently of their origin or dynamics, must be intrinsically related to patterns in the planar velocities. The mean velocities in the $X$-$Y$ projection around the Sun seen first in the \Gaia DR2 data by \citet{Katz2018} may be related to spiral arms and, interestingly, different types of arms could lead to different velocity patterns \citep[e.g.][]{Grand2015,Antoja2016}. Some of the observed diagonal ridges in the $V_\phi$-$R$ projection  \citep{Antoja2018,Kawata2018} and EDR3 data \citep{Antoja2021} have been tentatively related to transient spiral structure \citep{Hunt2018,Hunt2019,Khanna2019}, to resonances of density wave arms \citep{Hunt2019,Barros2020,Michtchenko2018}, or to recent crossings with the spiral arms \citep{Quillen2018}. Other ridges are very likely related to the bar \citep{Antoja2018,Hunt2018,Fragkoudi2019,Monari2019,Fragkoudi2020,Laporte2020}. Ridges can also appear due to an interaction with an external satellite \citep[e.g.][]{Laporte2019, Khanna2019}. 

Most likely, different processes are at play in the dynamics of the MW disc. \citet{Ramos2018} found different slopes in the ridges seen in the $V_\phi$-$R$ projection, which could indicate different origins. \citet{Friske2019} discovered a complex 
wave of $V_R$ as a function of
angular momentum and speculate that it could be a superposition of the effects of the bar and the spiral structure. It is, therefore, of the utmost importance to understand the dynamics of different types of spiral arms and identify observable signatures that help us distinguish them from each other and from other disturbances. In particular, the dynamics of tidally induced arms in the context of the MW remains poorly explored. 
 
 Here we study the effects of a perturbing satellite (similar to Sagittarius) on the 
 MW disc, paying special attention to how these structures are seen in phase space in the shape of ridges and other kinematic disturbances similar to those detected in the data. Our aim is to study their dynamics, distinguish them from other types of arms, and ultimately constrain the spiral structure of the MW and the interaction with Sagittarius. We first use simple toy models of impulsive and distant encounters to explore the formation and dynamics of spiral structure (\s\ref{s_simple}). We link the spiral arms generated in these models to the ridges in the $V_\phi$-$R$ projection and to a wave of $V_R$ as a function of angular momentum (\s\ref{s_rid}). Since we see that the frequency of these features becomes larger with time due to the phase-wrapping, we develop a method to recover the time of impact and the parameters of the potential through Fourier analysis (\s\ref{s_fre}). We then look at the more realistic N-body simulation by \citet{Laporte2018} that we compare against our simpler models (\s\ref{s_nbod}). 
 Finally, we compare models and real MW data from \Gaia DR3 \citep{Vallenari2022} in \s\ref{s_data}. We discuss our results in \s\ref{s_dis} and conclude in \s\ref{s_con}.


\section{Toy models of kinematic waves}\label{s_simple}

In this section we present a series of models of the Galactic spiral arms after an impulsive and distant tidal encounter, in particular, those often referred to as kinematic spiral waves. To introduce basic equations and give a brief historical overview, we start from the idea of the rotated ellipses (Sect.~\ref{s_ell}) and continue with models from 
orbit integrations for different distribution functions (Sect.~\ref{s_int}). These are all two-dimensional models (2D) that we compare to the more complex and three-dimensional (3D) N-body model of \s\ref{s_nbod}. In Table \ref{t_mod} we list all these models and specify their details. 

We use cylindrical coordinates ($R$, $\phi$, $V_R$, $V_\phi$). In all our models, the plots are presented with rotation in the clockwise direction and  
$\phi$ and $V_\phi$ are positive in this sense. The origin of $\phi$ is set at the negative $X$ axis.

For convenience, in the toy models presented below we use spherical power-law models that in the 2D idealisation are equivalent to an axisymmetric potential with orbits confined to the plane. In these potentials, the circular velocity curve is:
\begin{equation}\label{e_vc}
V_c(R)=V_{0} \bigl{(\frac{R}{R_0}\bigr)}^n,
\end{equation} 
where $n$ is the slope of the curve, and $V_{0}$ its value at $R=R_0$. We explore models with a perfectly flat circular velocity curve (logarithmic potential) and slightly decreasing ($n=-0.1$)  or increasing ($n=0.1$) circular velocity curves. The frequencies of nearly circular orbits (with guiding radius $R_g$) for models of the type of \e\ref{e_vc} are:
\begin{subequations}\label{e_fre}
\begin{alignat}{3}
\omega(R_g)&= \frac{V_0}{R_0^n}R_g^{n-1}   \\
\kappa(R_g)&=\frac{V_0}{R_0^n}\sqrt{2(n+1)}R_g^{n-1}   \\
\omega(R_g)-\frac{1}{2}\kappa(R_g)&=\frac{V_0}{R_0^n}\left(1-\frac{1}{2}\sqrt{2(n+1)}\right)R_g^{n-1}.  
\end{alignat}
\end{subequations}

\begin{figure}
   \centering
   \includegraphics[width=1\columnwidth]{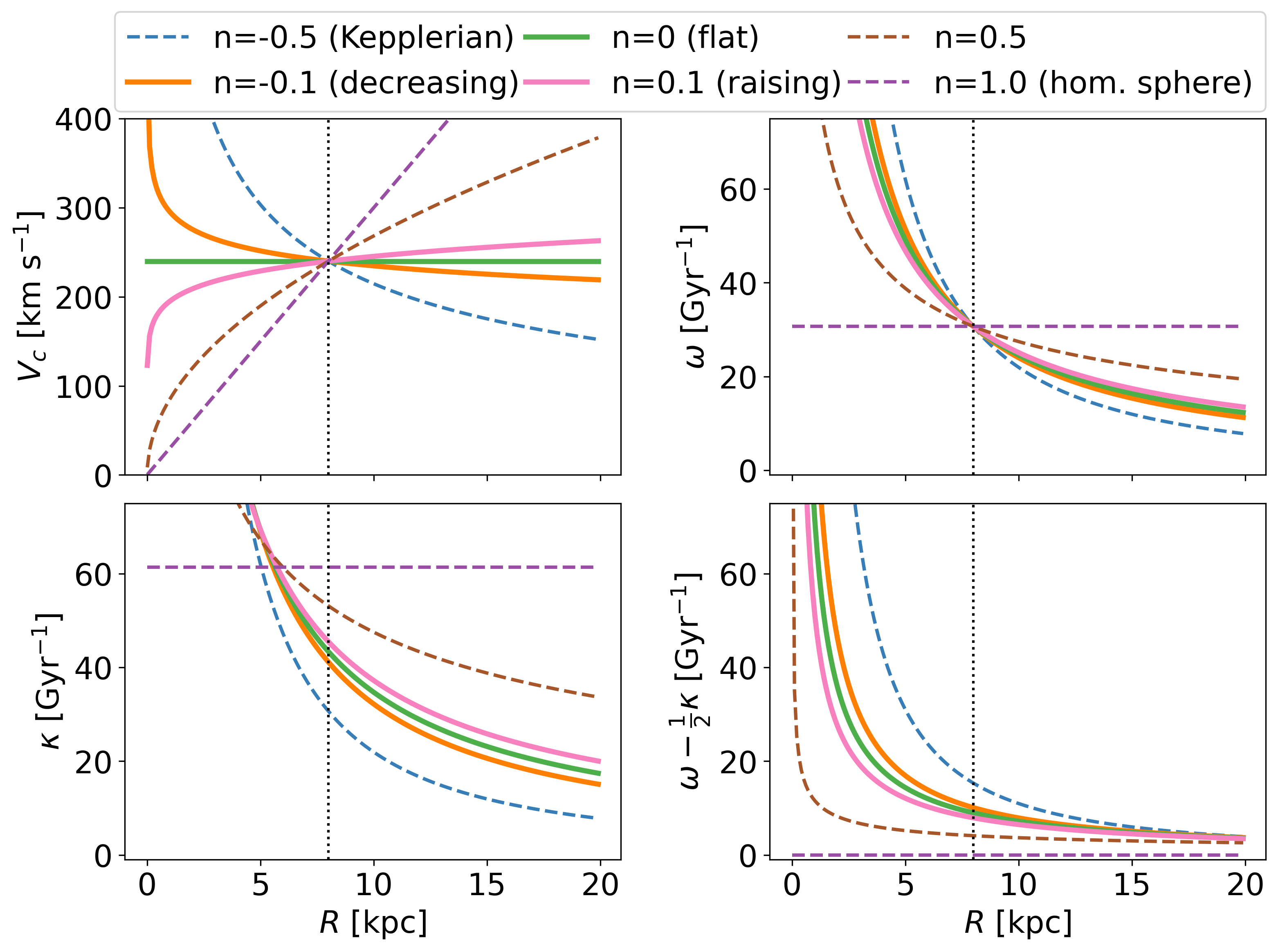}
   \caption{Circular velocity curve and frequencies for power-law models. We show potentials with different slopes indicated in the legend. The models used in the present study are shown as thicker solid lines, while dashed lines show other models for comparison. We show the circular velocity curve (top left), the azimuthal frequency (top right), the radial epicyclic frequency (bottom left) and \ok2 (bottom right). The vertical dotted line marks the Sun's position $R_0=8\,\kpc$.}
\label{f_freq}%
\end{figure}

The circular velocity curves as well as the azimuthal $\omega$, radial $\kappa$, and $\omega-\frac{1}{2}\kappa$ frequencies 
of these models are shown in \f\ref{f_freq}. For completeness we add other slopes of the circular velocity curves corresponding to n=0.5, -0.5 (Keplerian), and 1 (homogeneous sphere). As default values we use $V_0=240\,\kms$ and $R_0=8$ kpc, similar to the MW case \citep[e.g.][]{Gravity2021,Reid2020,Schonrich2010}. 

\begin{table}
\caption{Models used in this work. The columns indicate: the label of the model, the type of model, the potential used, the type of initial conditions, and the radial velocity dispersion at $R_0$ in units of $V_0$.}             
\label{t_mod}      
\centering                          
\begin{tabular}{l l l l l}        
\hline\hline                 
    Model & type & potential & IC&$\sigma$ [$V_0$] \\    
\hline                        
   0       & epicycles & $n=0$ & grid of orbits& 0. \\
   1cf   & orbits & $n=0$    &Dehnen, cold& 0.05 \\
   1hf    & orbits & $n=0$    &Dehnen, hot&  0.2\\
   1hd& orbits & $n=-0.1$ &Dehnen, hot& 0.2 \\ 
   1hr& orbits & $n=0.1$  &Dehnen, hot& 0.2 \\ 
   L18       & N-body &  &  & \\ 
\hline                                   
\end{tabular}
\end{table}

\subsection{Lindblad and Kalnajs ellipses}\label{s_ell}

The concept of the kinematic spiral waves was born around the  1950's, in the words of \citet{Lindblad1956}, as a ``line of thought which may lead to a satisfactory dynamical theory of the spiral phenomenon". The basis of the idea is that, in a reference frame that moves at frequency \ok2, orbits are seen as closed ovals, or equivalently, that the apsidal lines move at this frequency relative to each other if this frequency does not remain the same at all radii (this is the precessing rate of the closed ovals).

\begin{figure}
   \centering
   \includegraphics[width=0.9\columnwidth]{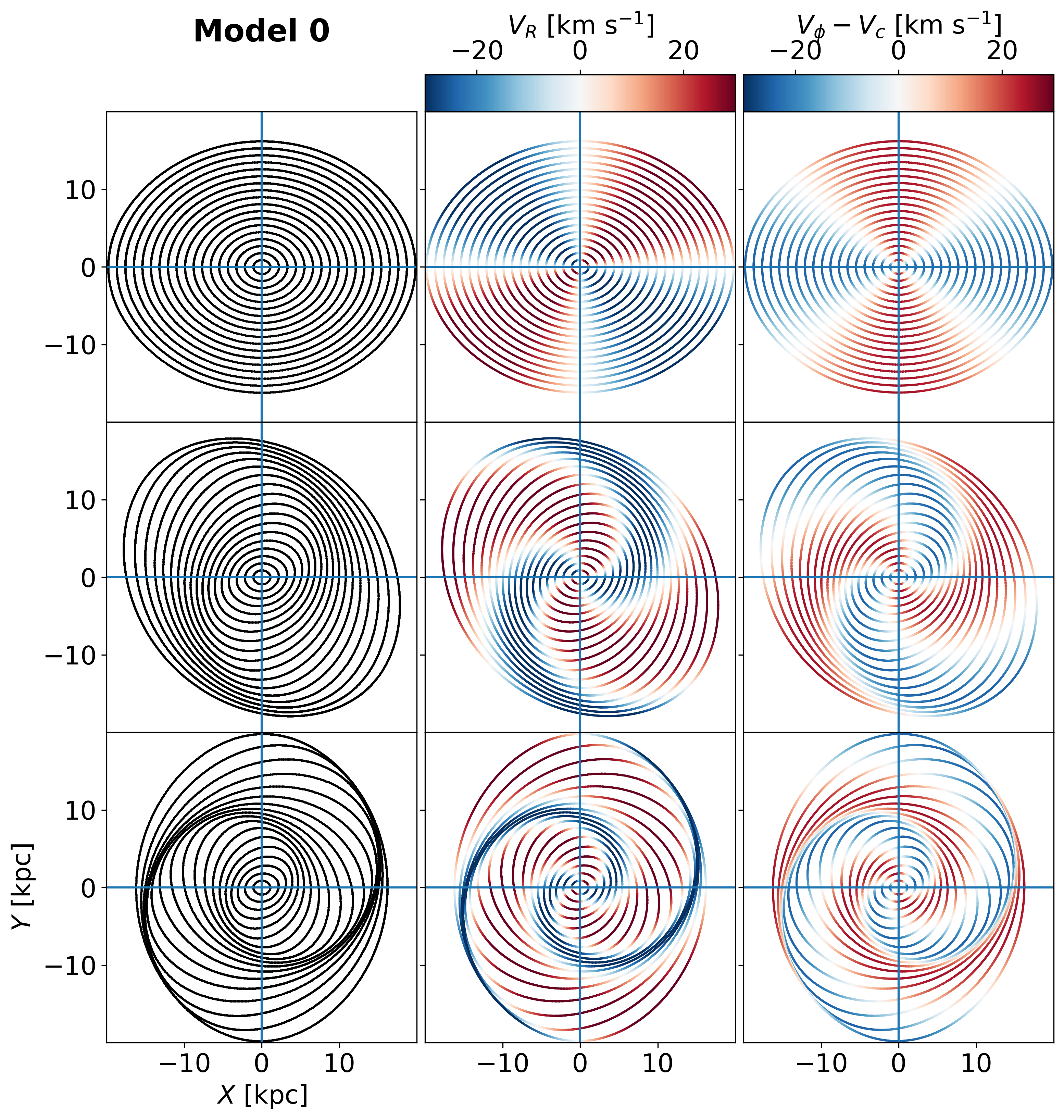}
   \caption{Rotated ellipses after \citet{Kalnajs1973}. The top panel shows a series of orbits in the reference frame of \ok2 for which the orbits close. In the first, second and third column, we plot the orbits, the orbits coloured as a function of radial velocity ($V_R$), and as a function of azimuthal velocity ($V_\phi-V_c$), respectively. The two bottom rows shows the same but for increasingly rotated ellipses.}
\label{f_ell}%
    \end{figure}

To illustrate this, we present our Model 0 in \f\ref{f_ell}. In this model we built a series of orbits using the equations of perfect epicyclic orbits:
\begin{subequations}\label{e_epi}
\begin{alignat}{4}
R&=R_g+AR_g\cos\left(\kappa(R_g) t+\psi_0\right)\\
    V_R&=-AR_g\kappa(R_g)\sin\bigl(\kappa(R_g) t+\psi_0\bigr)    \\
    \phi&=\phi_0+\omega(R_g) t-\gamma A\left(\sin(\kappa(R_g) t+\psi_0)-\sin(\psi_0)\right)\\
    V_\phi&=\frac{R_g^2 \omega(R_g)}{R}.
\end{alignat}
\end{subequations}
For simplicity, in this example we used a flat circular velocity curve ($n=0$). In these equations $A$ is the epicyclic amplitude and $t$ is the time. The constant $\psi_0$ gives the phase in the epicycle but is irrelevant in this particular exercise and it is set to 0. This is similar for $\phi_0$ which is the initial azimuth of the orbits and we took it to be 0. We sampled different $R_g$ and times $t$, choosing  $A=0.1$, and printed the orbits in the \ok2 reference system where they become closed orbits (\f\ref{f_ell}, top left). These ellipses will appear in this configuration only for a given instant of time as the different \ok2 at different radius would make the ovals precess at different rhythms.
\citet{Lindblad1956}, \citet{Lindblad1958} and \citet{Kalnajs1973} noticed, however, that this frequency can be nearly constant at intermediate and outer radius of disc galaxies (\f\ref{f_freq}). In this way, the set of  ovals remains quite fixed relative to each other  creating barred shapes, or, if the ovals appear rotated, persistent trailing spiral structure (left panel of the second row, which has been obtained by rotating the ellipses of the top panel giving them different values of $\phi_0$, in particular with the inner and outer ellipses differing by $150\deg$). We refer to \citet[][Sect. 6.2]{Binney2008} for more details.

In a more realistic case where  $\omega-\frac{1}{2}\kappa$ is not exactly constant with radius but has a small slope, a configuration like that of the top panels evolves towards more tightly wound spirality, such as in the second and third rows of \f\ref{f_ell} (the latter done by increasing the difference between inner and outer ellipses to $300\deg$). This structure slowly curls with time (at a rate of $\omega-\frac{1}{2}\kappa$), which is slower for raising circular velocity curves (right bottom panel of \f\ref{f_freq}), with the limit of perfect flatness (thus, no curling with time) for the homogeneous sphere ($n=1$).

The second and third columns of Fig~\ref{f_ell} indicate in colours the radial and azimuthal velocity (the later represented with respect to the circular velocity $V_c$), respectively, which follow from \es\ref{e_epi}. As \citet{Kalnajs1973} already noticed, the spiral arms form by the accumulation of orbits in-between their apocentres and pericentres, thus coinciding with regions where orbits are travelling inwards (negative $V_R$) and have $V_\phi-V_c=0$. Immediately past the arms, the orbits reach their pericentre and have, thus, maximum $V_\phi$ and $V_R=0$. The radial and azimuthal velocities show a $\pi/4$ rad shift in azimuth. 
With time, the regions of accumulations of orbits become more concentrated in configuration space, or, in other words, the angular region occupied by the spiral overdensitiies (and consequently the regions of $V_R<0$) is smaller compared to the inter-arm regions.

Interestingly, a configuration like the one in \f\ref{f_ell} (top) can result after a tidal interaction with a companion galaxy. For example, following \citet{Struck2011}, the gain in velocities for an impulsive and weak interaction is:
\begin{subequations}
\begin{alignat}{2}
    \Delta V_R&=\Delta V\frac{R\cos(2\phi)}{D}\\
    \Delta V_\phi&=-\Delta V\frac{R\sin(2\phi)}{D},
\end{alignat}\label{e_DV}
\end{subequations}
where $D$ is a scale parameter (the radius of the disc), and 
$\Delta V$ is a velocity amplitude obtained from the tidal constants and the ratio of the disc size to the distance of closest approach. 

Spiral patterns of the kinematic-wave type that roll up at a rate \ok2 have been reported in multiple N-body simulations \citep{Oh2008,Struck2011,Oh2015,Hu2018, Pettitt2016,Pettitt2018,Semczuk2017,Hu2018}. This idea has also been recently put in context of the MW by \citet{BlandHawthorn2021} who measured the windup frequency of the spiral structure formed in an N-body simulation following an impulsive impact of a Sagittarius-like perturber. Deviations from the $\omega-\frac{1}{2}\kappa$ rate are seen when a bar is also present and can depend on the self gravity of the disc and on time \citep{Pettitt2018,Oh2015}. In those cases, the frequency usually lies between \ok2 and \okfour. 
We see in \s\ref{s_nbod} that some of the pericentres of the N-body simulation that we explore also follow these kind of spiral patterns that slowly wind up with time.

\subsection{Model 1: Full orbital integration}\label{s_int}

\begin{figure}
   \centering
   \includegraphics[width=0.45\textwidth]{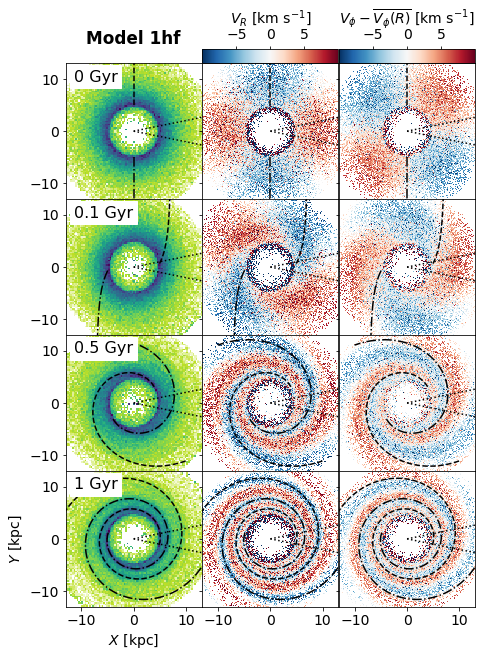}
   \caption{Tidal spiral arms of Model 1hf. We show the time evolution between 0 and 1 Gyr (indicated in the panels of the first column) of the disc in the $X$-$Y$ space, in density (left column), coloured by average radial velocity $V_R$ (middle) and average azimuthal velocity $V_\phi-\overline{V_\phi(R)}$ (i.e. with respect to the mean at the given radius, third column). The dashed and dashed-dotted lines show the prediction of the spiral arms location following the winding with \ok2. The sector of the disc marked with dotted lines is a region whose dynamics we examine later in \f\ref{f_r1}.}
\label{f_m1}%
    \end{figure}

In our modelling series that we dub Model 1, we used
{\it galpy}\footnote{\url{http://github.com/jobovy/galpy}} (\citealt{Bovy2015}) to integrate orbits. This is more realistic than our previous Model 0 in which all orbits followed perfect epicycles. We built four different models of this type (Table \ref{t_mod}) combining different initial conditions and potentials. We used two different sets of initial conditions from the distribution functions of \citet{Dehnen1999} with radial velocity dispersion at $R_0=8\,\kpc$ of 0.05 and 0.2 times $V_0$, which for $V_0=240\,\kms$ give 12 and 48 $\kms$, respectively. The labels of the models (Table \ref{t_mod}) indicate the dispersion with c and h standing for cold and hot, respectively. All of them have $200\,000$ test particles. We sampled the distribution only between 0.6 and 1.9 $R_0$ (4.8 and 15.2 kpc, respectively) in order to focus on the central parts of the disc. We use a scale-length of 1/4$R_0$, and an exponential radial-velocity-dispersion profile with scale-length $R_0$.  
We chose three analytical potentials from the power law family with $n=0$, $n=-0.1$ and $n=0.1$ (marked with solid thicker lines in \f\ref{f_freq}). In the model labels this slope is indicated with f, r, d indicating flat, rising and decreasing circular velocity curves, respectively. Although we generated distribution functions consistent with the corresponding slope, the initial conditions were integrated for 10 Gyr in order to diminish the effect of the phase-mixing patterns that appear as a consequence of not having a fully consistent distribution function. 

After this, we added the velocity kicks of \e\ref{e_DV} keeping positions unchanged. We used arbitrary values of $\Delta V=10\, \kms$ and $D=20\,\kpc$. These directly give the initial conditions for the orbital integration and are represented in the top row of \f\ref{f_m1} in the configuration space $X$-$Y$. The first column gives the positions of the toy orbits. Second and third columns indicate the average radial velocity $V_R$ and azimuthal $V_\phi-\overline{V_\phi(R)}$, respectively, where $\overline{V_\phi(R)}$ is the average azimuthal velocity at that radius. The quadrupole signature of \es\ref{e_DV} after the simulated satellite impact at $t=0$ is seen, with $V_R$ and $V_\phi$ shifted by a phase of $\pi/4$. With this scheme the orbits are distributed at special epicyclic phases depending on their location inside the disc.

The next rows of \f\ref{f_m1} show the evolution of the positions and velocities of  Model 1hf (with hot initial conditions and the logarithmic potential). The initial pattern rotates at an angular speed of \ok2, which depends on radius $R$, and thus its locus follows:
\begin{equation}\label{e_ok2}
    \phi(R,t)=\left(\omega(R)-\frac{1}{2}\kappa(R)\right)t+\phi_0.
\end{equation}
Because of the radial dependence of the rotation speed, the pattern winds up (shears) with time. In the top panel of \f\ref{f_m1} we show a dashed and a dot-dashed line corresponding to opposite azimuthal regions that have negative radial velocities at the initial times. In the following rows, these lines are wound up as described in \e\ref{e_ok2}. These lines mark the position of stars that are at an epicyclic phase of $\psi=\pi/2$ at each time. 
 A material arm (made of the same stars at all times)
would instead phase-mix at a rate of $\omega$, that is, with a much steeper dependence on $R$ and thus much faster. Another fundamental difference with material arms is that here the  pattern is composed of stars with common epicycle phase at a given time, and thus stars that instantaneously are on top of the crest of the wave change with time. 

Continuing with \f\ref{f_m1}, the epicyclic oscillations lead to regions of high orbital crowding, creating a trailing two-armed spiral structure. In particular, the location of the maximum density (locus of the arms) corresponds to regions with stars whose epicyclic phases passed apocentre and now move towards pericentre. This leads to the correspondence between the locus of the arms and regions with negative radial velocity. This also results in the regions of negative $V_R$ getting compressed in configuration space (i.e. their azimuthal width at a fixed radius gets smaller). The inter-arm regions correspond to positive radial velocities. This is the same behaviour as in the formalism presented in \s\ref{s_ell} (\f\ref{f_ell}). For this particular model, even after 1 Gyr, clear spiral structure is seen in the range of radii from 5-10 kpc (similar to that of the volume sampled by \Gaiaf).

\begin{figure}
   \centering
   \includegraphics[width=0.5\textwidth]{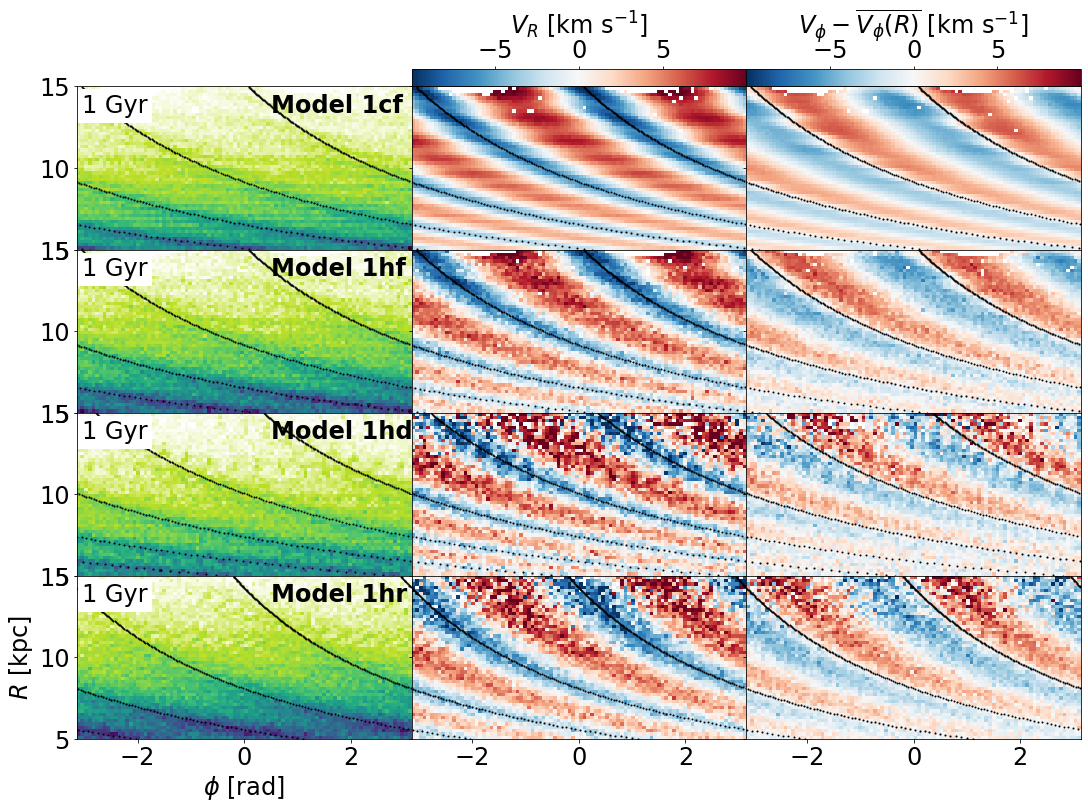}
 \caption{Spiral arms of different models in polar coordinates. We show the $\phi$-$R$ projection at 1 Gyr, in density (left column), coloured by average radial velocity $V_R$ (middle) and average $V_\phi-\overline{V_\phi(R)}$ (right). The black lines show the predicted location of the spiral arms.}
\label{f_polar}%
    \end{figure}
    
A polar representation often does a better job at showing the spiral arms in density. \ff\ref{f_polar} shows the positions and  velocities in the $\phi$-$R$ plane now for different models  at 1 Gyr. Again, the azimuthal component of the velocity corresponds broadly to a phase shift of $\pi/4$ with respect to $V_R$.
 The first two rows show different levels of kinematic temperature of the initial conditions from cold to hot (1cf,  1hf). The potential used was the same and, therefore, the winding rate and the exact location of the arms and associated velocity regions are equal. The differences are related to the level of dispersion (noise) in the average positions and velocities. 
 
We note that additional substructure appears both in density and velocities in \f\ref{f_polar}. For example, a clear pattern of small substructure in density, $V_R$ and $V_\phi$ is seen especially for the cold model (first row). These patterns must be the result of the phase mixing but their exact mathematical descriptions were not derived here where we focus on the structure of our interest (i.e. the spiral arms). We note that similar substructure is observed in  similar models such as from \citet{Struck2011}.

The three bottom rows of \f\ref{f_polar} compare different slopes of the circular velocity curve for the hotter initial conditions: Model 1hf with $n=0$ flat (second row), Model 1hd with $n=-0.1$ decreasing circular velocity curve (third row) and Model 1hr with $n=0.1$ rising circular velocity curve (last row). The different models have different slopes of the frequencies with radius, in particular of \ok2: as shown in \f\ref{f_freq} this frequency is flatter for the rising circular velocity curve, and becomes progressively steeper for the flat and the decreasing curves. This results in a more tightly wound pattern in the 1hd case (smaller pitch angle).

     \begin{figure}
   \centering
   \includegraphics[width=0.5\textwidth]{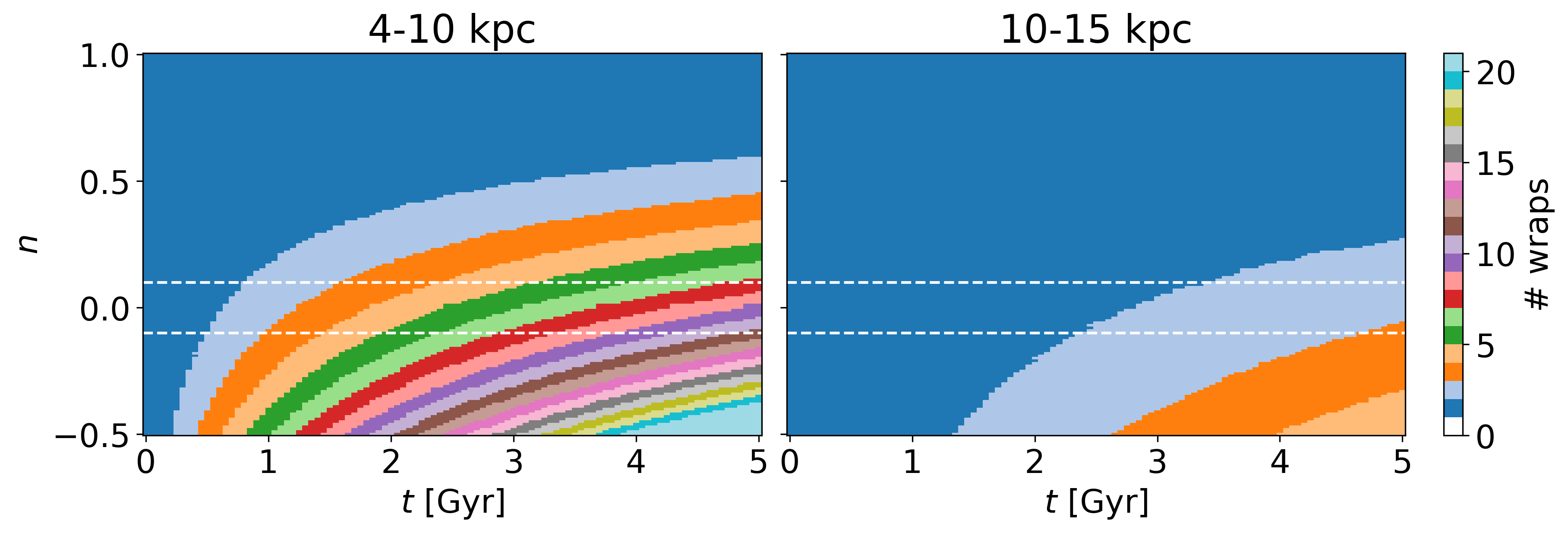}
   \caption{Number of spiral wraps. We show in colours the expected maximum number of spiral wraps at different radius ranges (4-10 and 10-15 kpc in the first and second columns, respectively) as a function of time $t$ and slope of the circular velocity curve $n$. We have used 21 for the upper limit of the colour bar but the number of wraps goes beyond this in the bottom part of the left panel.}
\label{f_wra}%
    \end{figure}

The degree of winding with time of these kinematic-wave toy models depends, therefore, both on the time (after impact) and on the slope of the circular velocity curve. For a certain time, the number of spiral wraps crossing at a certain azimuth $\phi_i$ within the range between $R_{min}$ and $R_{max}$ is given by $\lfloor\frac{\phi(R_{min})-\phi_i}{2\pi}\rfloor - \lfloor\frac{\phi(R_{max})-\phi_i}{2\pi}\rfloor$, where $\phi(R)$ is the one defined by \e\ref{e_ok2}. 
\ff\ref{f_wra} shows the number of spiral wraps at certain radius ranges as a function of time and of the slope of the circular velocity curve. Since this number depends on $\phi_i$, for the figure we show the maximum number of wraps encountered for all possible $\phi_i$. 
 At inner radius, the number of wraps is much larger at a given $t$ and $n$, as expected from the slope of the curve \ok2. At a certain time and radius, the number of wraps is larger the more negative the slope of  the circular velocity curve $n$ is. In the outer parts of this idealised disc, the number of wraps is small even for larger times after the supposed impact. 

Below we explore in detail these kinematic waves seen in other projections of phase space. This helps us to link them with the ridges in the $R$-$V_\phi$ plane and a $V_R$ wave with angular momentum, which are features observed in the MW.

\section{Ridges and angular momentum waves}\label{s_rid}

\begin{figure*}
   \centering
   \includegraphics[width=0.8\textwidth]{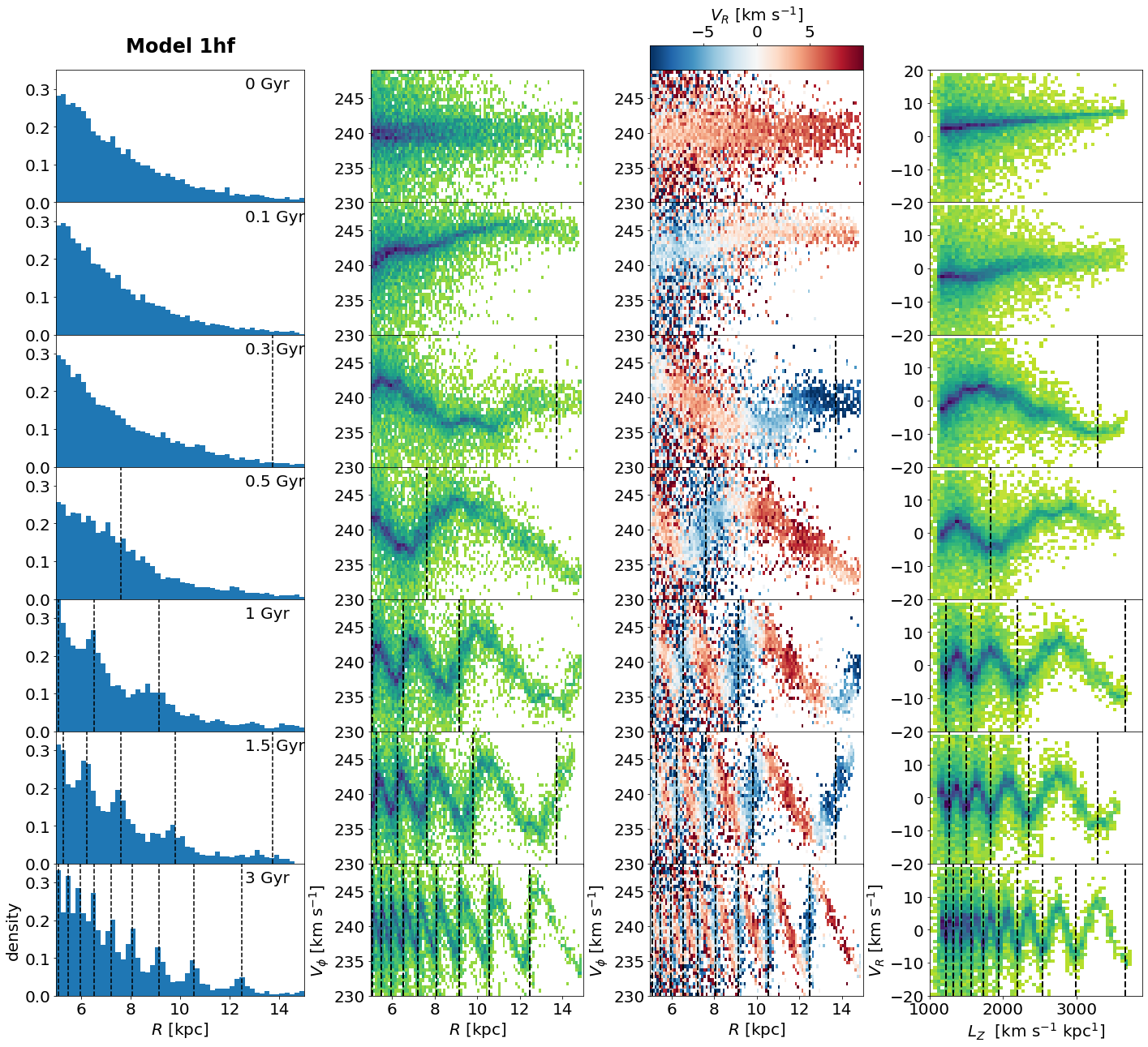}
   \caption{Ridges and waves for Model 1hf. We show the time evolution between 0 and 3 Gyr (indicated in the first column) of the disc region at $\phi=\pi\pm 0.2$ rad (marked with dotted lines in \f\ref{f_m1}). The first column shows a normalised density histogram of the radius of the particles $R$. The second and third columns show the $R$-$V_\phi$ projection in density and coloured by mean $V_R$, respectively. The fourth column shows the particles in the $L_Z$-$V_R$ projection. The vertical dashed lines in all panels indicate the prediction of the spiral arm locations following a winding rate of \ok2. }
\label{f_r1}%
    \end{figure*}

In \f\ref{f_r1} we focus on a particular azimuthal region of the disc to look at other projections of phase space. We start with Model 1hf (hot initial conditions and flat circular velocity curve).
We take only the region of  $\phi=[\pi-0.2, \pi+0.2]$ rad (marked with dotted lines in \f\ref{f_m1}), which corresponds to a sector of $\approx$22.9 deg centred on the positive $X$ axis. In the first column we show a density histogram of the radius of the particles. We see some clear bumps that correspond to locations of higher density, that is to say the spiral arms generated in the model. The vertical dashed black lines show the predicted positions of the arms. These are obtained by first inverting \e\ref{e_ok2} and using \e\ref{e_fre}c to obtain: 

\begin{equation}\label{e_R}
    R(\phi,t)={\left[\frac{V_0}{R_0^n} \frac{\left(1-\frac{1}{2}\sqrt{2(n+1)}\right) t}{(\phi-\phi_0) } \right]}^{\frac{1}{1-n}}.
\end{equation}
Then we find the positions of the multiple wraps, $R_m$ ($m=1,2,etc$) by adding $\pi$ to the azimuth $\phi$ an integer number of times $m$, that is changing $\phi$ to $\phi+m\pi$ to obtain $R_m=R(\phi+m\pi,t)$. For this particular example we take $\phi_0=\pi/2$ (the initial azimuth of the arms in our model) and the azimuth $\phi=\pi$ (the centre of the sector chosen).

As shown before, 
the number of wraps of spiral arms crossing the region becomes larger with time. The spatial separation between wraps at a given time increases with $R$ due to the smaller slope of \ok2 with $R$ at large radii.

The second column of \f\ref{f_r1} shows the density in the $R$-$V_\phi$ space, while the third one is the same but coloured by average $V_R$. At the beginning ($t=0$, first row), there is a flat distribution of $V_\phi$ as a function of $R$ because in this model the circular velocity curve is flat. The dispersion in $V_\phi$ is larger for inner radii, as expected for typical disc galaxies and exactly as generated in the initial conditions. The $\phi=\pi\pm0.2$ rad region corresponds to initially positive radial velocities (thus dominated by red colours in the third column) as can be seen also in the top middle panel of \f\ref{f_m1}. As time goes by, sawtooth-shaped oscillations 
appear in the $R$-$V_\phi$ projection, with the parts of positive slope roughly corresponding to negative radial velocities (blue colours) and the decreasing parts of the oscillations having positive radial velocities (red colours). As we have seen before, the negative $V_R$ regions are linked to the kinematic spiral arms and thus we see that in this projection the spiral arms correspond to the blue bands. We also plot vertical dashed black lines in the radius where the arms are located, which indeed coincide with the centres of the blue bands. The blue and red velocity bands tend to get stepper with time, with the blue ones becoming almost vertical due to the spiral arms becoming more tightly wound (smaller pitch angle). 
The bands of alternating negative and positive radial velocity also become closer due to the winding of the pattern.

\begin{figure*}
   \centering
   \includegraphics[width=0.8\textwidth]{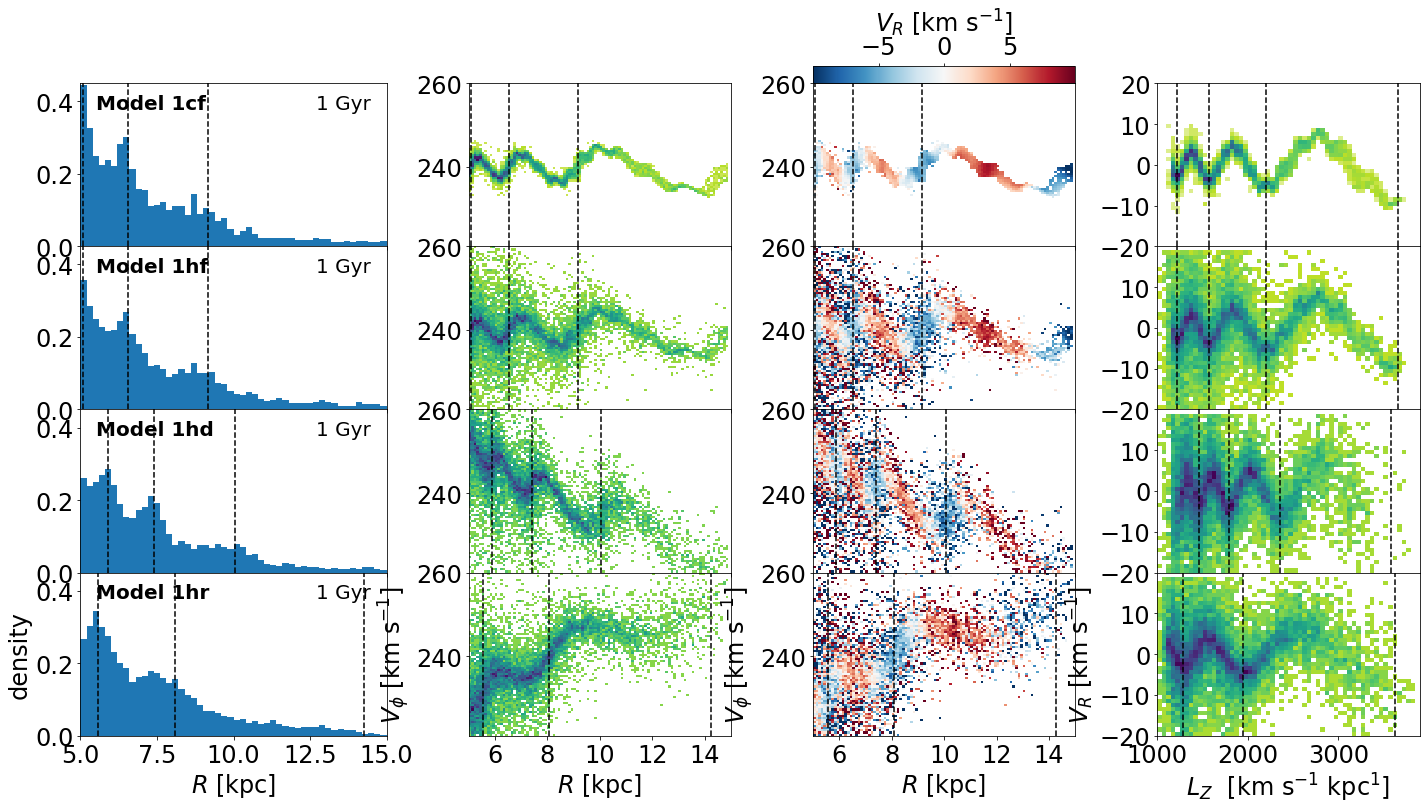}
   \caption{Ridges and waves for different models. The columns are the same as in \f\ref{f_r1} but now each row shows a different model for the time of 1 Gyr.}
\label{f_rall}%
\end{figure*}

Finally, in the fourth column of \f\ref{f_r1}, we show the radial velocity as a function of the angular momentum. The initial distribution shows simply a straight line with a small positive slope, resulting from the imposed initial perturbation that increases linearly with $R$ -- combined with the transformation from $R$ to $L_Z $ for a perfectly flat circular velocity curve in an azimuth where particles have not gained or lost angular momentum (panel in the third column of the first row in \f\ref{f_m1}). Again we see larger velocity dispersion at inner radii. With time, oscillations in $V_R$ appear, with increasingly larger frequency. The growing amplitude with $L_Z$ comes from the imposed velocities at the beginning (\e\ref{e_DV}). 
The black dashed vertical lines correspond to the location of the arms in the space of $L_{Z}$ (corresponding to the minima in $V_R$). They were obtained similarly as $R_m$. We first transformed \e\ref{e_R} to angular momentum (taking into account that the arms are made of orbits exactly at their guiding radius)  
 $L_{Z} (\phi,t)=R(\phi,t)V_c(R(\phi,t))$:
\begin{equation}\label{e_L}
    L_{Z}(\phi,t)=\left(\frac{V_0}{R_0^n}\right)^{\frac{2}{1-n}}{\left[ \frac{\left(1-\frac{1}{2}\sqrt{2(n+1)}\right) t}{(\phi-\phi_0)} \right]}^{\frac{1+n}{1-n}},
\end{equation}
and then set $\phi_0=\pi/2$, $\phi=\pi$, and changed $\phi$ to $\phi+m\pi$, that is computing $L_{Z,m}=L_{Z}(\phi+m\pi,t)$.
For a flat circular velocity curve ($n=0$), \e\ref{e_L} becomes:
\begin{equation}\label{e_Lf}
    L_{Z}(\phi,t)={V_0}^2\frac{\left(1-\frac{1}{\sqrt{2}}\right)t} {\phi-\phi_0}.
\end{equation}

At a given time, the frequency of the $V_R$ oscillation depends on $L_Z$, as it also depends on $R$. This can be seen in \e\ref{e_Lf} since the separation between valleys of minimum $V_R$ is $\Delta L_Z\equiv L_{Z,m}- L_{Z,m+1}\propto \frac{1}{m^2+k_1m+k_2}$, where $k_1$ and $k_2$ are constants and $m$ takes larger values for the valleys at smaller $L_Z$ 
 (and similarly for $n\neq0$ with \e\ref{e_L}). However, from \e\ref{e_L} one also realises that the oscillations in $V_R$ have constant period if plotted as a function of $L_Z^{-1}$ instead of $L_Z$. 
 We note a similar behaviour for non-flat circular velocity curves as a function of $L_Z^{\frac{n-1}{n+1}}$. 
 This is explored further in \s\ref{s_fre}.

\ff\ref{f_rall} shows the ridges and waves for the different flavours of Model 1 at 1 Gyr of evolution. In the first two rows we compare the different initial conditions but use the same logarithmic potential model. The effects of the velocity dispersion are observed between panels. 
Taking the second, third and fourth rows we can compare different slopes of the circular velocity curves ($n=0, -0.1, 0.1$). The effect of $n$ is to modify the overall slope of the $V_\phi$-$R$ curve making it flat, decreasing, and rising. We also note less ridges and less oscillations for the rising curve model (1hr) due to the slower winding rate, as already noticed in  \f\ref{f_polar}. The ridges in the third column appear more conspicuous and elongated in Model 1hd since the $V_R$ pattern, which follows lines of negative slope in this diagram, is favoured by the decreasing circular velocity curve. In fact, the $V_R$ pattern in the third column follows lines of constant angular momentum (in accordance with what is seen in the fourth column). On the contrary, the slope of the ridges in the $V_\phi$-$R$ (second column) changes with the slope of the circular velocity curve combined with the sawtooth pattern. As for the wave of $V_R$ with $L_Z$, we note that the wavelength varies between models with different $n$.

\begin{figure}
   \centering
   \includegraphics[width=0.4\textwidth]{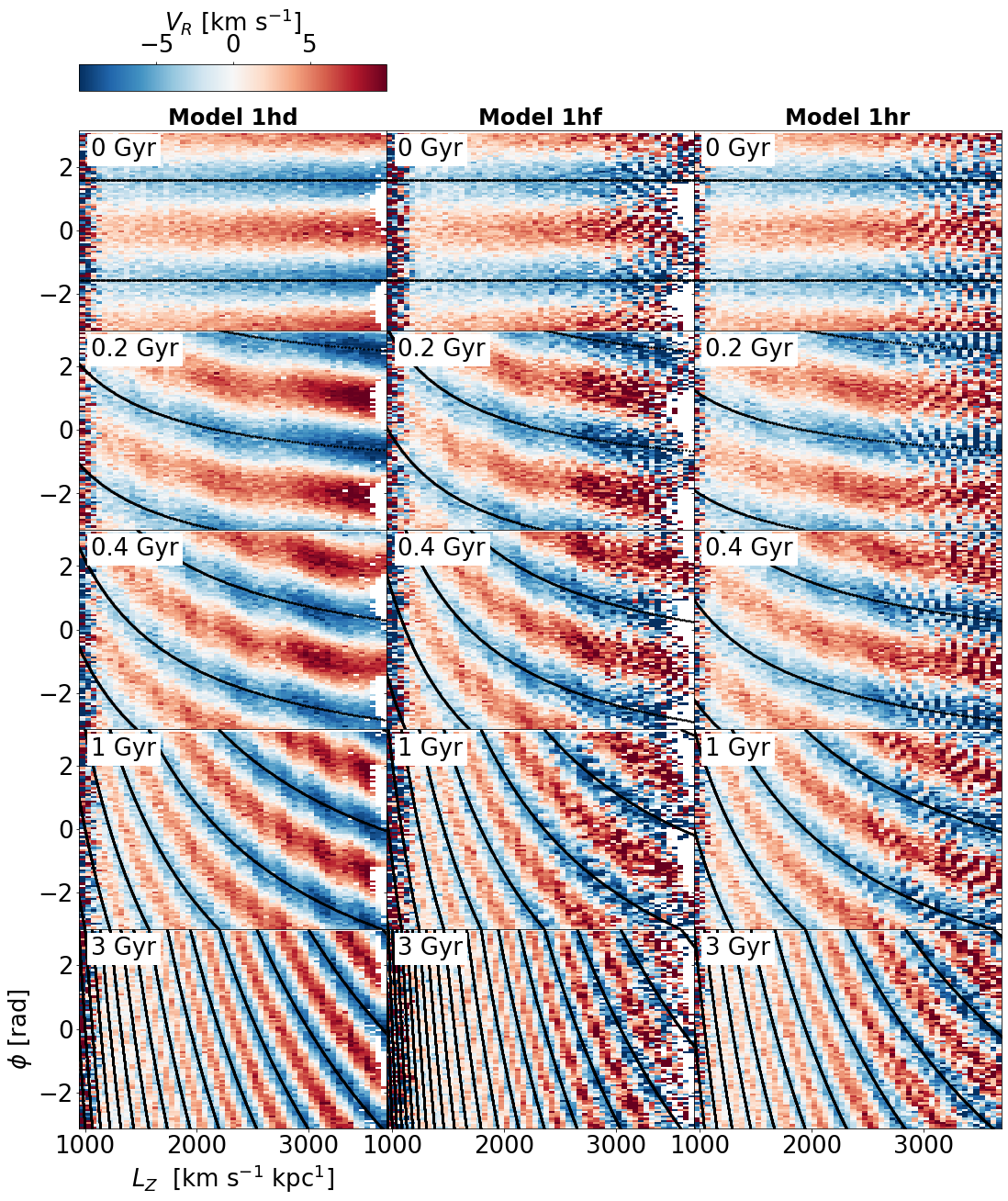}
   \caption{Azimuthal dependency of the $L_Z$-$V_R$ wave. We show the waves by plotting the mean $V_R$ as a function of azimuth $\phi$ and $L_Z$ for different models (from left to right) and their time evolution from 0 to 3 Gyr (from top to bottom).}
\label{f_LZ}%
    \end{figure}

To examine the azimuthal dependency of the ridges, 
in \f\ref{f_LZ} we show the radial velocity as a function of $L_Z$ and $\phi$ for different models with different circular velocity curves (columns) and times (rows). As already explained, for the kinematic spiral arms, the regions of negative (blue) $V_R$ correspond to the arms. The black lines show the locations of the arms in this projection. They were obtained by computing the minima of the $V_R$ wave following \e\ref{e_L}, where in practice one needs to plot $L_Z$ as a function of $\phi\bmod\pi$. In the first row ($t=0$), there is the perfect quadrupolar alternation of negative and positive $V_R$ regions for all models as imposed in the initial conditions. As time goes on, more wraps of the spiral are seen in the shape of inclined bands. As already seen, the number of wraps (bands) depends both on the time and the slope of the circular velocity curve. In the same manner, the (negative) slope of the blue and red pattern increases as a function of time and decreases with $L_Z$ at a fixed time.

\begin{figure}
   \centering
   \includegraphics[width=0.45\textwidth]{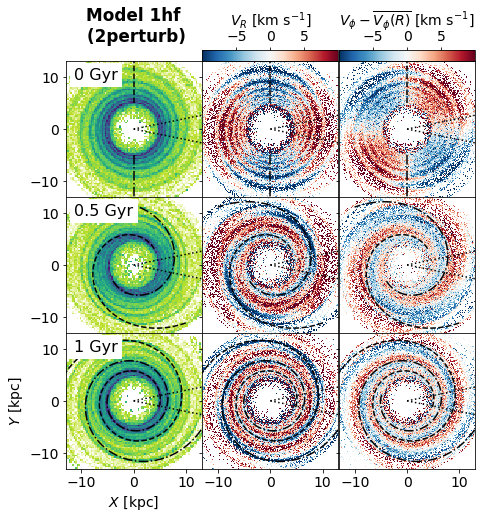}
   \caption{Spiral arms of Model 1hf for two consecutive perturbations. The first perturbation starts at $t=-2\,\Gyr$ and has $\Delta V_1=10\,\kms$, and the second starts at $t=0$ and has $\Delta V_2=15\,\kms$. The plot has the same structure as \f\ref{f_m1} but fewer times are shown. The lines show the arms location consequence of the second impact.}
\label{f_m2per}%
    \end{figure}

        \begin{figure*}
   \centering
   \includegraphics[width=0.8\textwidth]{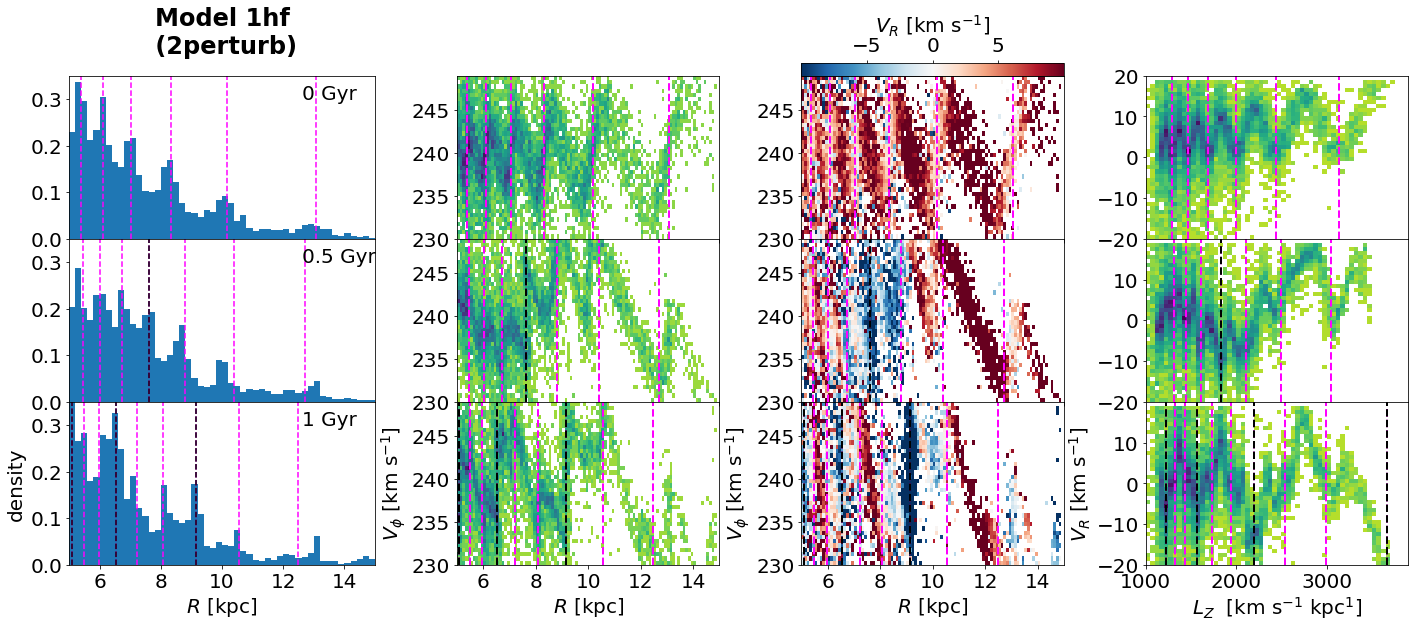}
   \caption{Ridges and waves for Model 1hf for two consecutive perturbations. The plot has the same structure as \f\ref{f_r1} but fewer time steps are shown. The pink and black dashed vertical lines are the location of spiral arm wraps from the first and second perturbation, respectively.)
}
\label{f_r1per}%
    \end{figure*}
    
As a final analysis with these simple models, we simulated 
two consecutive impacts which would reflect a more general situation during a satellite interaction with a disc. To do this, our initial conditions are those of Model 1hf after 2 Gyr of the first impact. At that time the particles receive a new velocity kick following \e\ref{e_DV} with $\Delta V_2=15\,\kms$, thus higher than the previous one (which was of $\Delta V_1=10\,\kms$). The initial conditions and the configuration at 0.5 and 1 Gyr of evolution after the second impact are shown in \f\ref{f_m2per}. A series of largely wrapped spirals remaining from the first impact are observed at the initial stage to which the new pattern of \e\ref{e_DV} is superposed. At 0.5 Gyr a combination of the two patterns is visible, although the later one dominates visually due to its larger wavelength and amplitude. The inspection of the azimuthal sector at $\phi=\pi$ (\f\ref{f_r1per}) shows that the superposition of the two patterns is visible in the ridges and in the angular momentum space as two superposed waves with different wavelengths.

The inspection of the figures in this section indicate that the wavelength of the oscillations in various projections give information on both the time after perturbation and the circular velocity curve. We even see that signatures of multiple impacts may persist as superposed waves with different wavelengths. We therefore explore this further in \s\ref{s_fre}, using Fourier analysis.
    
\section{Frequency analysis}\label{s_fre}

 In this section we perform Fourier transforms (FT) of the signatures of the kinematic spiral arms in phase space to estimate the time of impact (i.e. the start of the phase mixing process) using the fact that they  increase their frequency with time.

\begin{figure}
   \centering
   \includegraphics[width=0.5\textwidth]{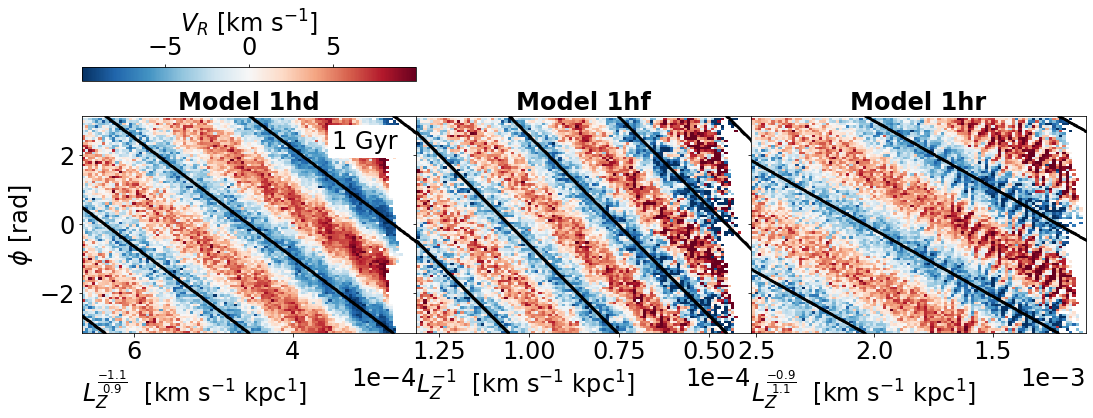}
   \caption{Azimuthal dependency of the $L_Z$-$V_R$ wave in the transformed $L_Z$ space. We plot the mean $V_R$ as a function of azimuth $\phi$ and $L_Z^{\frac{n-1}{n+1}}$ for different models (different columns) and at a time of 1 Gyr. Compared to \f\ref{f_LZ}, the transformation from $L_Z$ to $L_Z^{\frac{n-1}{n+1}}$  makes the colour bands straight and equispaced, improving the FT analysis and allowing a determination of the time of impact.}
\label{f_Lt}%
    \end{figure}

As already noticed, at a given time the frequency decreases with $L_Z$, making the frequency of this signal ill-defined. Motivated by the shape of the curves of minimum $V_R$ in the $L_Z$-$\phi$ space (\e\ref{e_L}) we can transform the $L_Z$ coordinate to make it linearly dependent on $\phi$:
\begin{equation}\label{e_invL}
L_Z^{\frac{n-1}{n+1}}=\left(\frac{V_0}{{R_0}^n}\right)^{\frac{-2}{1+n}}\frac{1}{1-\frac{1}{2}\sqrt{2(n+1)}}\frac{\phi-\phi_0}{t}.
\end{equation}
In this way, the curves of minimum $V_R$ are straight and equally spaced in the $L_Z^{\frac{n-1}{n+1}}$-$\phi$ space. For the particular case of $n=0$ in a flat circular velocity curve, this means simply using $L_Z^{-1}$:
\begin{equation}\label{e_invL0}
L_Z^{-1}=\frac{V_0^{-2}}{1-\frac{1}{\sqrt{2}}}\frac{\phi-\phi_0}{t}.
\end{equation}
As an example, in \f\ref{f_Lt} we show the different flavours of Model 1 at time $t=1$ Gyr in the transformed space $L_Z^{\frac{n-1}{n+1}}$-$\phi$. These depict straight and equally spaced bands. The black lines correspond to the analytical description given in \e\ref{e_invL}. If a certain azimuthal coverage is available in the data, a fit of \e\ref{e_invL} to the blue bands gives, for known $V_0$, $n$ and $R_0$, a direct determination of $\phi_0$ and $t$ from the slope and intercept. In practice one could also do a 2D FT for this fit. However, since our real data (see \s\ref{s_data}) do not cover a wide range of $\phi$ we explore how to gain (partial) information from data with limited azimuthal coverage.

    \begin{figure}
   \centering
   \includegraphics[width=0.35\textwidth]{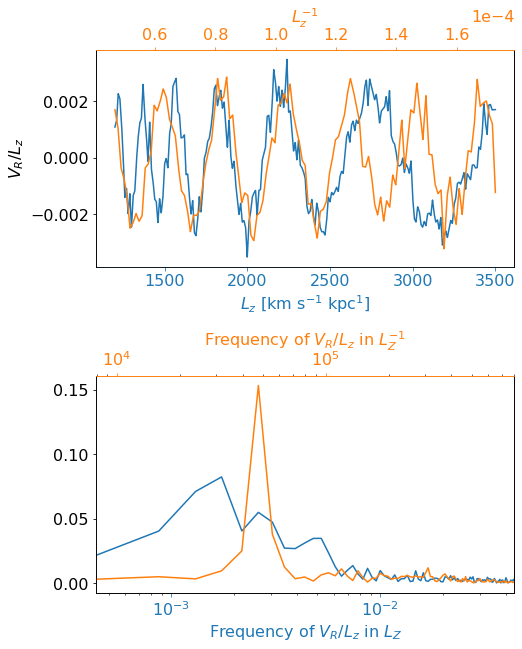}
   \caption{Example of frequency analysis of the $V_R$ wave for the Model 1hf at time $t=2$ Gyr. Top: The blue line is the average $V_R$ as a function of $L_Z$, while the orange line is in the scale of $L_Z^{-1}$, shown in the top horizontal axis. Bottom: Fourier amplitude as a function of frequency in $L_Z$ (blue and bottom axis) and  in  $L_Z^{-1}$ (orange, top axis). The frequencies are in units of ${(\kmskpc)}^{-1}$ (blue) and  $\kmskpc$ (orange).}
\label{f_f1}%
    \end{figure} 
 
From \f\ref{f_Lt} and from \e\ref{e_invL} we can see that in a certain azimuthal location ($\phi$) the valleys of minimum $V_R$ are now equispaced a distance (i.e. the wavelength):
\begin{equation}\label{e_Lsep}
\Delta L_Z^{\frac{n-1}{n+1}}=\left(\frac{V_0}{R_0^n}\right)^{\frac{-2}{1+n}}\frac{1}{1-\frac{1}{2}\sqrt{2(n+1)}}\frac{\pi}{t}.
\end{equation}
Therefore, with this equation we can try to recover the time of impact $t$ in our simple models by measuring the frequency of the oscillations in $V_R$ with a FT in this transformed coordinate. 

We start by an example with Model 1hf in a given azimuth that we take as $\phi=\pi$ corresponding to the one in \f\ref{f_r1}, and first choose the time $t=2$ Gyr. The signal ($V_R$ oscillation) and FT (amplitude against frequency) are shown in the top and bottom panels of \f\ref{f_f1}, respectively. We use $V_R/L_Z$ to obtain a signal with constant amplitude with $L_Z$ (countering the imposed initial conditions, \e\ref{e_DV}), although this does not affect the FT. For comparison, we do the FT of the $L_Z$-$V_R$ curve (blue colours and bottom horizontal axis) and the case of the transformed coordinate $L_Z^{\frac{n-1}{n+1}}$-$V_R$, that for this case of $n=0$ is $L_Z^{-1}$-$V_R$ (orange curves and top horizontal axis). In the FT in the $L_Z$ coordinate we see multiple peaks over a wide range of frequencies. This is due to the fact that the frequency is not constant along the $L_Z$ axis. When we use $L_Z^{-1}$, however, the signal shows a uniform frequency and the FT reveals a clear peak.

    \begin{figure*}
   \centering
   \includegraphics[width=0.9\textwidth]{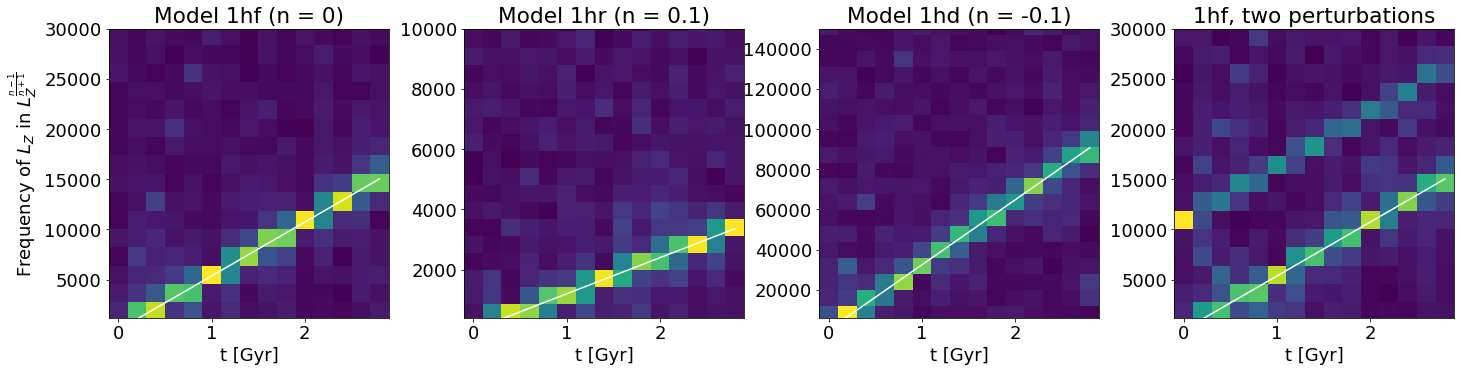}
   \caption{Frequency analysis of the $V_R$ wave for different models. From left to right we show Models 1hf, 1hr, 1hd, and 1hf with 2 perturbations. The colour indicates the amplitude of the Fourier transform as a function of time when the $L_Z$ is transformed to $L_Z^{\frac{n-1}{n+1}}$. The diagonal lines in all models show the increase in frequency with time, which is faster for the decreasing circular velocity curve mode (1hd). In the ideal model with two perturbations, two lines of the frequency corresponding to the two different impacts are detected at all times.}
\label{f_fmodels}%
    \end{figure*}
    
\ff\ref{f_fmodels} (left) shows the amplitude of the FT now for all times for the same Model 1hf (colour map). A clear diagonal band is observed indicating an increase in the frequency with time following the expected $\propto t$ from the inverse of \e\ref{e_Lsep}. The analytical formula is superposed in white and we see a perfect match between the predicted and the computed frequency, meaning that the FT analysis allows to recover the impact time for this simple model. In the two following right panels in \f\ref{f_fmodels} we show the computed frequency for the models with different slope of the circular velocity curve $n$. As expected, models with circular velocity curves with positive (negative) slope have a slower (faster) rhythm of phase mixing. We note that to do the FT in the transformed space $L_Z^{\frac{n-1}{n+1}}$, we have assumed that we know $n$. Alternatively, one could search for the $n$ that produces a clearer FT peak.  

Finally, in the right panel of \f\ref{f_fmodels} we show the FT amplitude of Model 1hf with two superposed perturbations offset by 2 Gyr (see Fig. \ref{f_r1per}). For this idealised model, two lines corresponding to the signal of each individual perturbation are identified at all times, from which we can recover both impact times using \e\ref{e_Lsep}.

In this section we have built simple models in which the dynamics can be described analytically.  
We now explore the more realistic N-body model using the simulation of \citet{Laporte2018} (\s\ref{s_nbod}) and continue with an analysis of the MW data (\s\ref{s_data}).

\section{N-body model}\label{s_nbod}

Our last model is the simulation L2 by \citet{Laporte2018}. Briefly, this idealised live N-body simulation followed the interaction of a Sgr-like object with a MW-like host from the time of virial radius crossing to the present-day. The simulation was run with the Gadget-3 code \citep{Springel2005b}. The host dark halo and bulge were represented by \citet{Hernquist1990} profiles with masses  $M_{h}=10^{12}\,\rm{M_{\odot}}$ and $M_{b}=10^{10}\,\rm{M_{\odot}}$, and scale radius of $a=53\,\kpc$ and $a=0.7\,\kpc$, respectively. The disc followed an exponential profile in the radial direction and a $\rm{sech}^{2}$ profile in the vertical one with scale radius $R_{d}=3.5\,\kpc$ and scale height $z_{d}=0.53\,\kpc$, and a total stellar mass of $M_{d}=6\times10^{10}\,\rm{M_{\odot}}$. The model was generated with GalIC \citep{Yurin2014} which was modified to include adiabatic contraction following the \citep{Blumenthal1986} implementation. The Sgr-like progenitor was modelled as a Hernquist sphere of $8\times10^{10}\,\rm{M_{\odot}}$ and $a=8\,\kpc$, made to match a NFW halo of virial mass $6\times10^{10}\,\rm{M_{\odot}}$ with concentration $c=28$. This corresponds to a concentration that is about twice as concentrated as the mean of the \citet{Gao2008} mass-concentration relation which has about 0.3 dex scatter. For more details we refer the reader to \citep{Laporte2018}.

\begin{figure}
   \centering
   \includegraphics[width=0.4\textwidth]{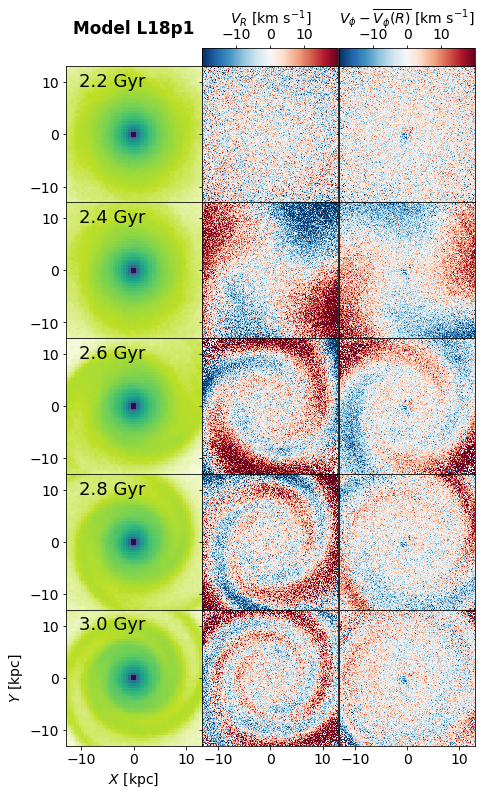}
   \caption{Spiral arms of Model L18p1. This is the same as \f\ref{f_m1} but for the L18 model immediately before the first  pericentre (2.4 Gyr).}
\label{f_m3a}%
    \end{figure}

\begin{figure}
   \centering
   \includegraphics[width=0.4\textwidth]{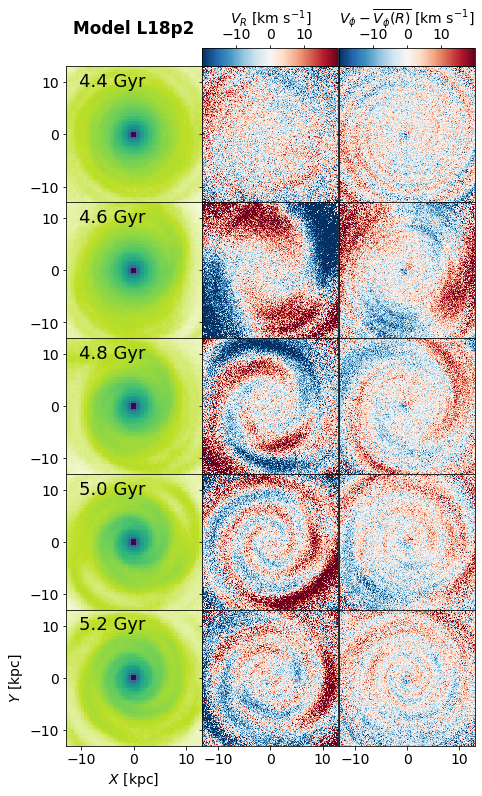}
   \caption{Spiral arms of Model L18p2. This is the same as \f\ref{f_m3a} but for immediately before the second pericentre (4.5 Gyr).}
\label{f_m3b}%
    \end{figure}

\begin{figure}
   \centering
   \includegraphics[width=0.4\textwidth]{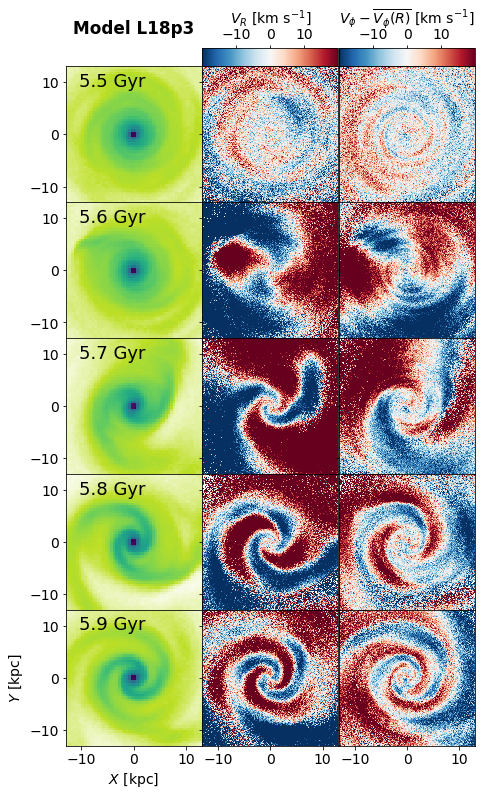}
   \caption{Spiral arms of Model L18p3. This is the same as \f\ref{f_m3a} but for immediately before the third  pericentre (5.6 Gyr).}
\label{f_m3c}%
    \end{figure}

\begin{figure}
   \centering
   \includegraphics[width=0.4\textwidth]{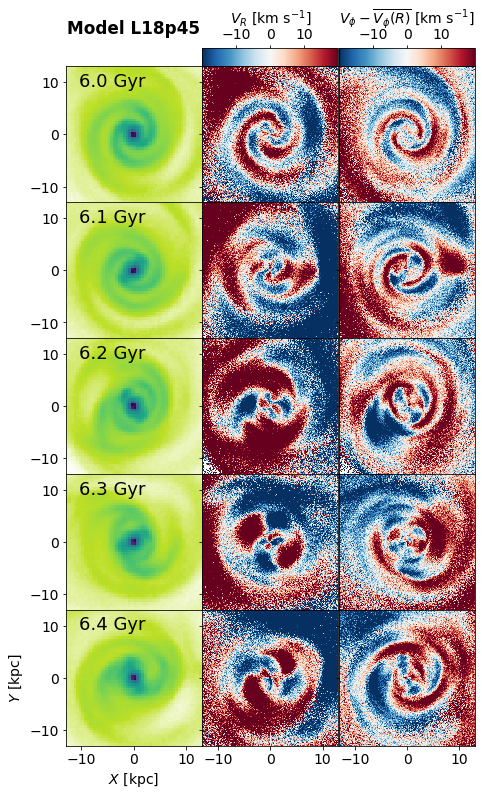}
   \caption{Spiral arms of Model L18p45. This is the same as \f\ref{f_m3a} but for the fourth  pericentre (6.1 Gyr). The successive percientre at 6.4 occur also in this set of times.}
\label{f_m3d}%
    \end{figure}

The pericentre passages of Sagittarius occur approximately at 2.3, 4.5, 5.6, 6.2 and 6.4 Gyr after the start of the simulation (the last pericentre roughly corresponds to the present time).  We present the analysis of this model in four different time intervals starting always at a time  previous to a pericentre: 1) the first pericentre (Model L18p1), 2) the second one (Model L18p2), 3) the third one (Model L18p3), and 4) the final sequence of snapshots when the last two pericentres (fourth and fifth) occur (Model L18p45). 
There are notable differences between these four time intervals. 
First, we examine the disc structure seen face-on and the velocity moments in \fs\ref{f_m3a}, \ref{f_m3b}, \ref{f_m3c}, and \ref{f_m3d}, corresponding to the different time ranges.

Immediately after the first pericentre (panel at 2.4 Gyr in \f\ref{f_m3a}) there is a situation very similar to our simpler models: a quadrupole pattern in velocity (second and third columns) closely following the behaviour described by \e\ref{e_DV} (but already slightly coiled at 2.4 Gyr). Indeed, as shown in Fig. 11 of \citet{GrionFilho2021} where the same model is explored, there is a strong $m=2$ mode in the Fourier decomposition of the radial and azimuthal accelerations at 2.2 Gyr (together with an $m=1$ mode) for particles at radius of 10 kpc. With time, the velocity patterns wrap and a bi-symmetric spiral structure in density appears (first column). As before, the spiral arms correspond to regions of negative radial velocity  and null azimuthal velocity with respect to the azimuthal mean. A dipole in $V_\phi$ appears in the very central kpc which might be due to small centring uncertainties. 

For the next time interval, in the first row of \f\ref{f_m3b}, we see that at 4.4 Gyr (that is 2 Gyr after the first pericentre) the spiral wraps from the first pericentre are still visible both in density and in velocities. After the second pericentre (panel at 4.6 Gyr),  we again see a dominant $m=2$ mode in the velocities that once more starts winding and forming a two-armed spiral structure. The impact of this second pericentre is slightly stronger than the first one, reflected in larger kick velocities (for example comparing panels at 2.4 and 4.6 Gyr). The signatures of the small scale ripples from the previous pericentre are hard to distinguish soon after the next impact (hardly visible by 0.3 Gyr after the second impact; middle row of \f\ref{f_m3b}). However, the signal of the previous impact can be seen better in other phase space projections (see below).

After the third pericentre (\f\ref{f_m3c}), a similar generation of a two-armed pattern is observed in density and in the velocity moments. In this case, though, the initial velocity perturbations follow \e\ref{e_DV} (quadrupole signature) only  in the inner disc but not in the outer parts. This is due to the perturber crossing at a much closer distance (in fact it crosses the disc) and thus the condition of a distant impact is only roughly fulfilled in the inner parts. The resulting inner spiral, especially in velocities, is not as symmetric as before but in general evolves as in previous pericentres. The spiral structure in this case is even stronger than previously (both in density and velocities), as expected after a closer pericentre. Again some ripples stemming from the previous (second) impact are visible at the first times after the third pericentre. The outer parts of the disc follow slightly different patterns in velocity.

The time span in \f\ref{f_m3d} includes the last two pericentres of this model. These occur at separations shorter than the dynamical times of the stars in the disc outskirts \citep{GrionFilho2021}, producing large impact on the disc dynamics. The density structure of these final stages of the simulation is more complex, with two spiral arms but also rings and a central bar with ansae (appearing at about 6.1 Gyr). The velocity patterns associated to the spiral arms are not so clear, perhaps masked by the signal of the bar. The inner quadruple of the bar is obvious and the quadrupole with a phase shift of $\pi/2$ in some snapshots (e.g. 6.3 Gyr) could be due to the surpassing of corotation \citep{Muhlbauer2003} or the start of two new tidal arms given that these structures appear somehow sheared in the following snapshot. 

\begin{figure*}
   \centering
   \includegraphics[width=0.7\textwidth]{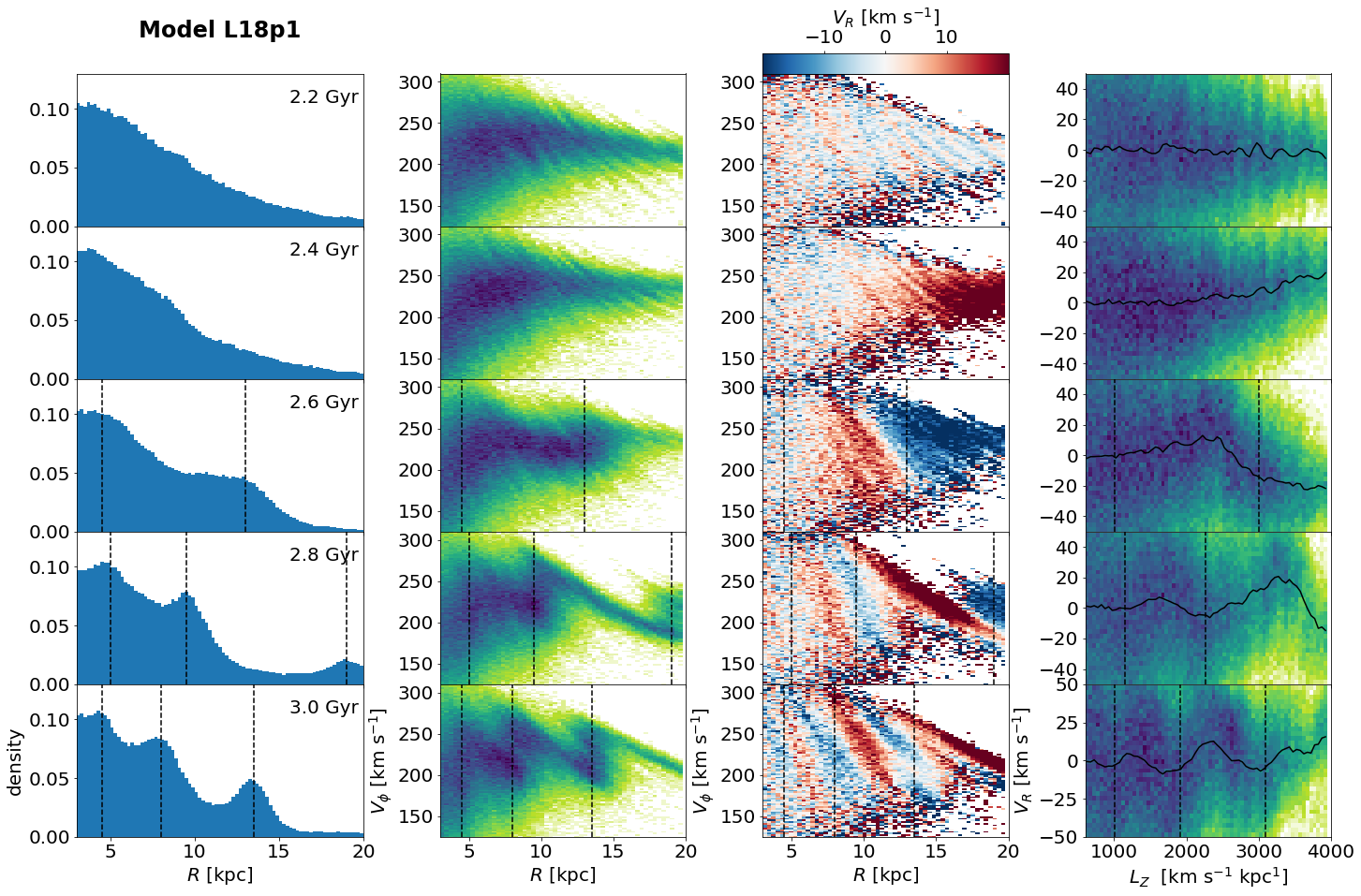}
   \caption{Ridges and waves for Model L18p1. The structure of the panels is as for \f\ref{f_r1}. The average $V_R$ in bins of $L_Z$ is shown as a black line in the last column.}
\label{f_r3a}%
    \end{figure*}
        
\begin{figure*}
   \centering
   \includegraphics[width=0.7\textwidth]{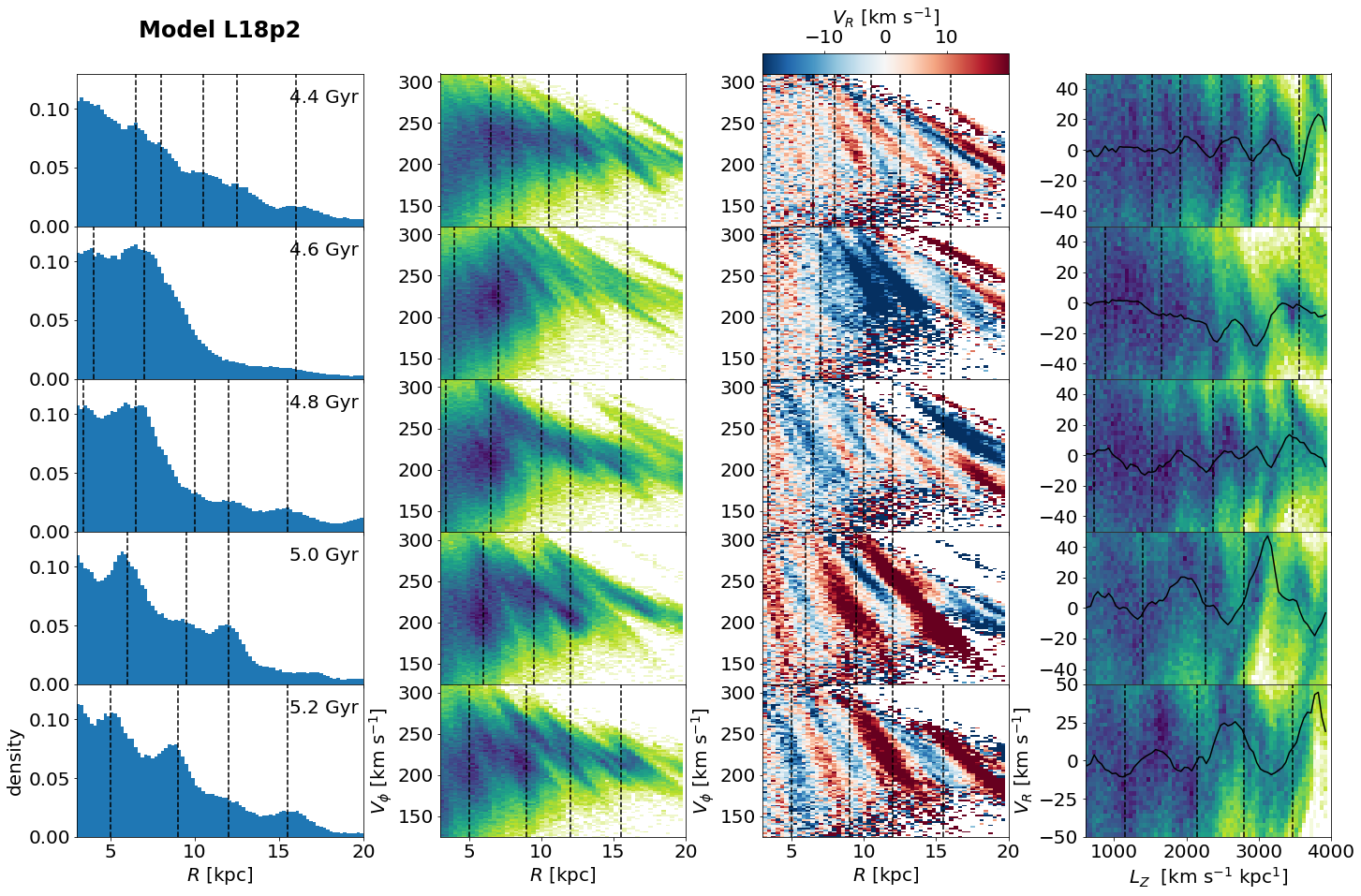}
   \caption{Ridges and waves for Model L18p2. The structure of the panels is as for \f\ref{f_r1}.}
\label{f_r3b}%
    \end{figure*}    
\begin{figure*}
   \centering
   \includegraphics[width=0.7\textwidth]{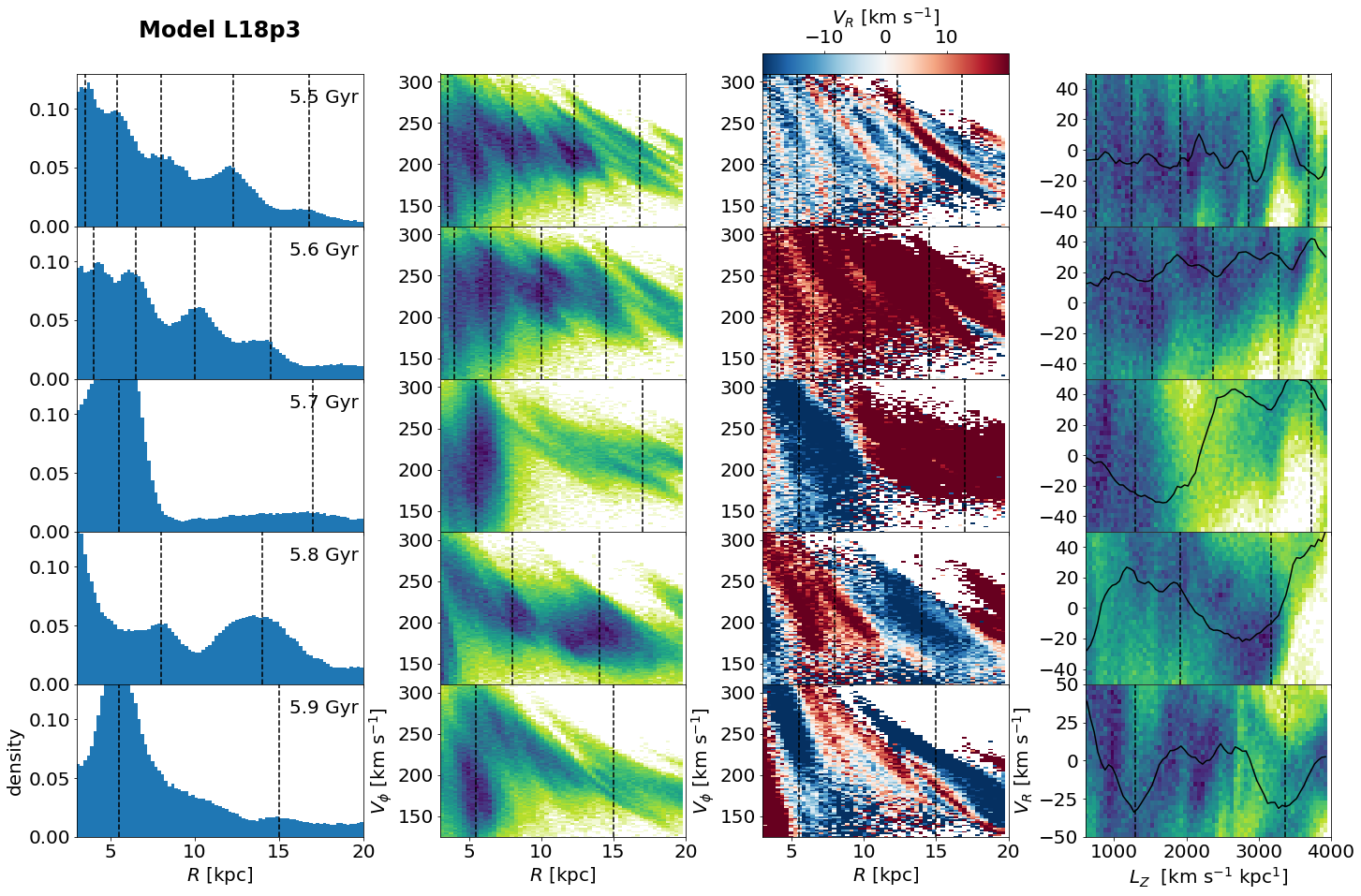}
   \caption{Ridges and waves for Model L18p3. The structure of the panels is as for \f\ref{f_r1}.}
\label{f_r3c}%
    \end{figure*}    
\begin{figure*}
   \centering
   \includegraphics[width=0.7\textwidth]{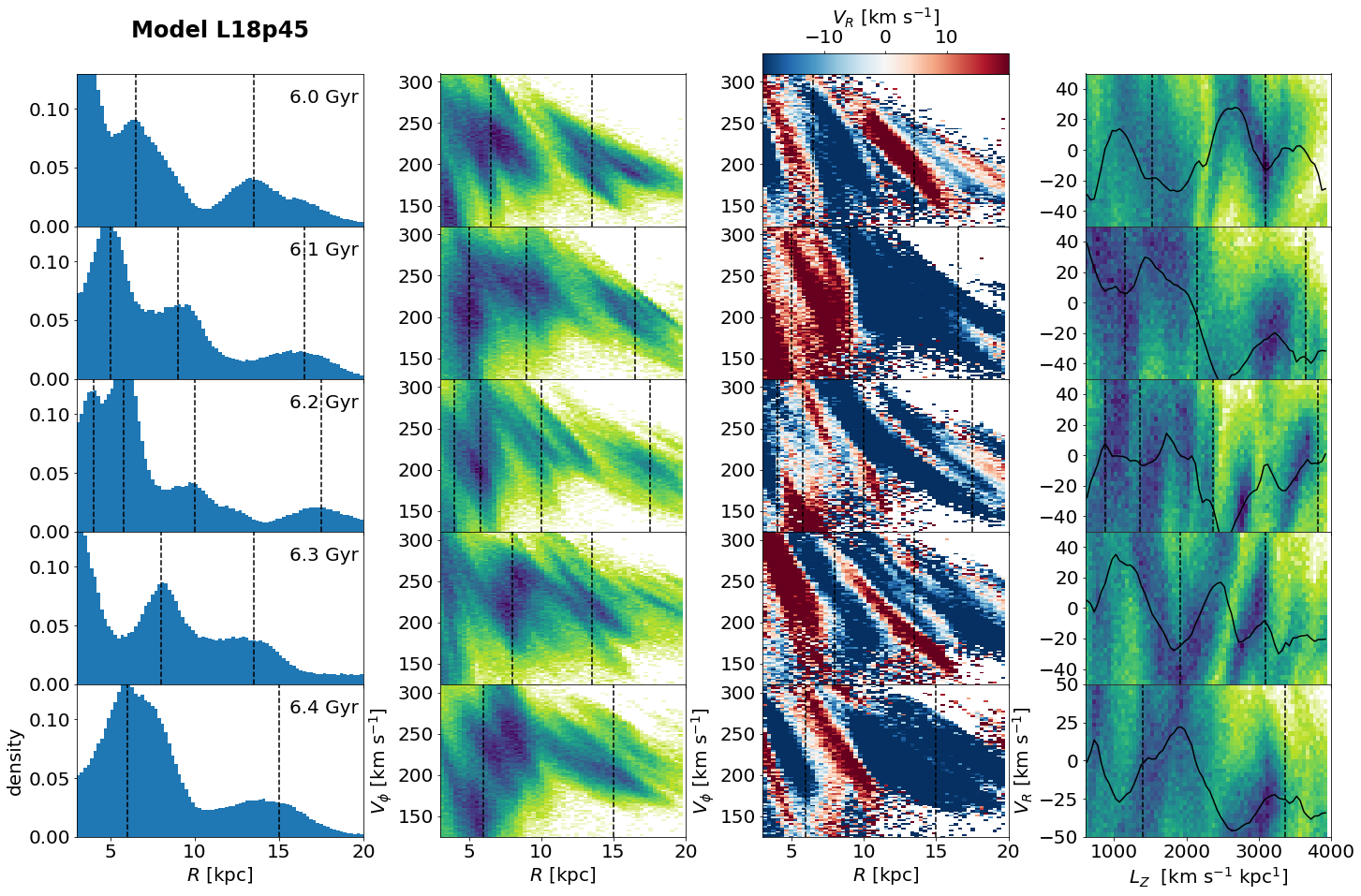}
   \caption{Ridges and waves for Model L18p45. The structure of the panels is as for \f\ref{f_r1}.}
\label{f_r3d}%
    \end{figure*}
    
Now we examine the dynamics corresponding to a certain azimuthal angle in the disc  ($\phi=\pi\pm0.2$) examining the four different time ranges in \f\ref{f_r3a}, \ref{f_r3b}, \ref{f_r3c} and \ref{f_r3d}, as we did for the simple models (\f\ref{f_r1}, \ref{f_rall} and \ref{f_r1per}). For this model, we could not use \e\ref{e_R} and thus in these figures we visually located the position of the overdensities in the histogram (first column) that correspond to the arm locations (or the bar in some cases) and marked them with vertical lines. For the vertical lines in the $L_Z$-$V_R$ projections (fourth column) again we could not use \e\ref{e_L} but we estimated the location of the minima by doing $L_Z=RV_c$ as in the analytical model but with $V_c$ obtained from the N-body model, in particular, at a time previous to the perturber.

\begin{figure*}
   \centering
   \includegraphics[width=0.75\textwidth]{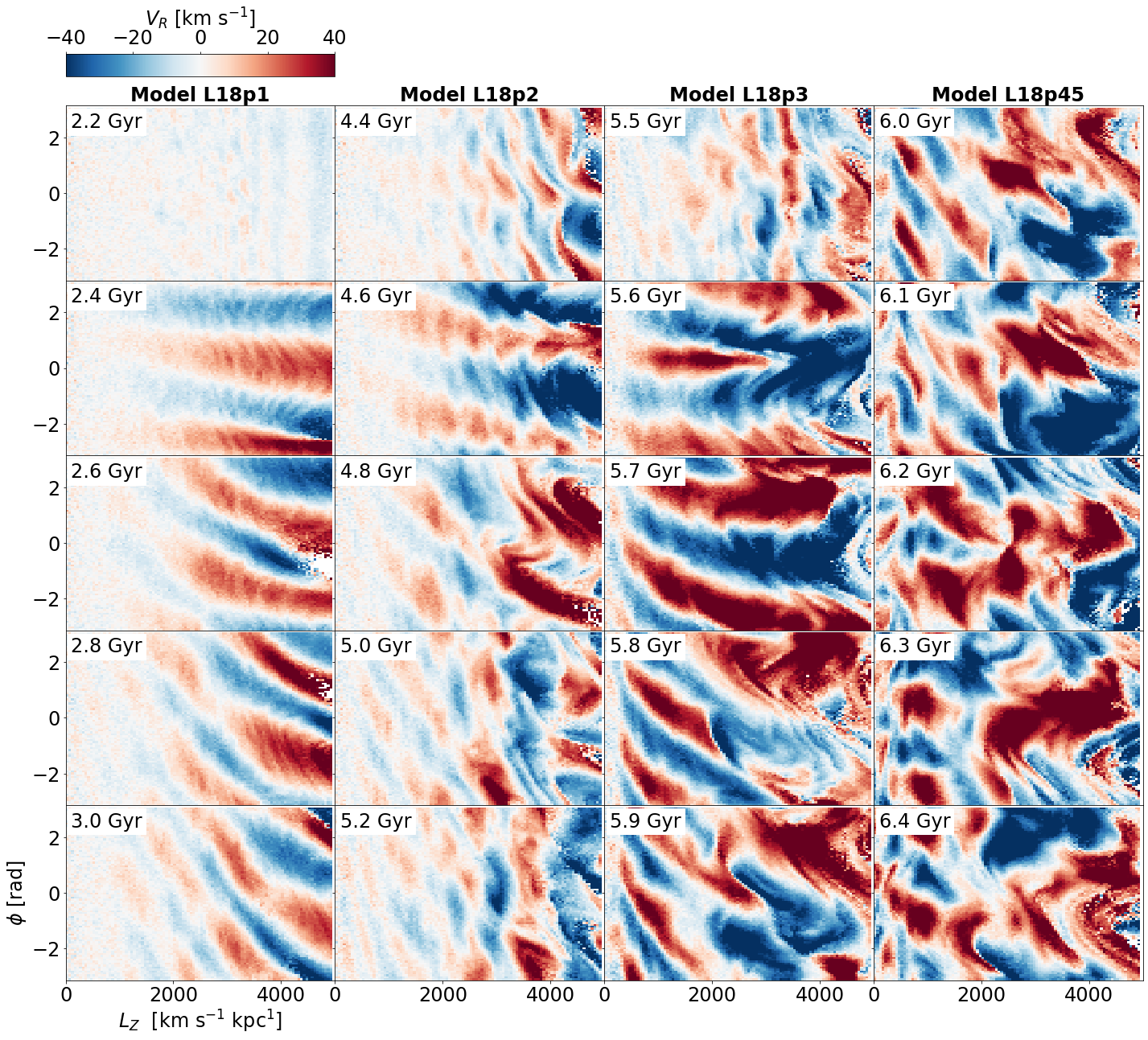}
   \caption{Azimuthal dependency of the $L_Z$-$V_R$ waves for the different time ranges of Model L18 organised in columns.}
\label{f_LZVR3}%
    \end{figure*}

Starting with the first pericentre (L18p1), in \f\ref{f_r3a} we observe ridges and an $V_R$ wave with a very similar behaviour to Models 1. This is not surprising since we already saw similarity in the $X$-$Y$ projection. Ridges form in a sawtooth shape (second column), with the ascending parts corresponding to the spiral arms in density (marked with vertical lines) and negative radial velocity, and positive radial velocities in the interarm regions and the descending ridges. We again see alternation of positive and negative radial velocities in the third column\footnote{Before the first pericentre at 2.2 Gyr we see some diagonal ridges indicating phase mixing originating from the initial conditions of the simulation. At the time where the perturber approaches for the first time, these structure are relatively well-mixed with no further consequences in the period dominated by the perturber. These patterns are also recognizable in \f\ref{f_LZVR3}.\label{f_1}}. By 3 Gyr, three spiral arm wraps are seen in the histogram of radius (first column) with three associated ridges in the middle columns. A wave of $V_R$ in $L_Z$ forms with a progressively shorter wavelength with time similar to Models 1. The vertical lines in this projection correspond almost perfectly to the minima, as in our simple models.

The situation after the second pericentre (\f\ref{f_r3b}) is reminiscent of our simple model with two consecutive pericentres (\f\ref{f_r1per}) but with the added complexity in the L18 model. Even-though we saw that the small spiral ripples in the $X$-$Y$ projection that remained from the first pericentre seemed to disappear in a relatively short time, here we keep seeing the corresponding ridges: at 4.6 Gyr just after the new pericentre occurs, the velocity $V_R$ seems overall dominated by negative values (blue) but we still see some small scale pattern of blue-red alternate (as evidenced also by the wave in $L_Z$-$V_R$ in the fourth column). By 4.8 Gyr, for example, one clear bump in the histogram of $R$ is seen corresponding to the new spiral wrap at this azimuth (clearly seen in \f\ref{f_m3b}) but we observe ridges with substructure in the second column and multiple blue and red bands in the third column. In these panels some of the thinner ridges do not follow the sawtooth shape anymore. This is because the two patterns of ridges (from the current and previous pericentre) superpose. 

Overall the $L_Z$-$V_R$ wave in \f\ref{f_r3b} follows the same behaviour as in the earlier pericentre but with  additional higher frequency undulations.  The effects of the first pericentre are thus still visible after 3 Gyr after the first impact (at 5.2 Gyr, bottom panels).  Although we suspect that this substructure also exist in configuration space, it is much harder to see due to the statistical noise. As in other work in different context, the velocity substructure seems to be preserved much better \citep{Helmi1999b}.
Finally, the vertical lines in the fourth column coincide only approximately with the minima of the wave. This could be because in this case it is more difficult to locate the spiral arms in the first column due to the interference between the two perturbations and because the use of the $V_c$ from early times is not accurate anymore. The dynamics of the third pericentre (\f\ref{f_r3c}) also shows  waves of different wavelengths from the different pericentres, similarly to L18p2. 

In the final 2 pericentres (\f\ref{f_r3d}) there are clear bumps in the $R$ histograms (associated to the spiral structure) that correlate with the thicker up-going ridges in the second column and with  the negative radial velocities bands in the third column (and valleys in the fourth column). However, the correspondence between vertical lines and minima is good only in some cases possibly for the same reasons stated above. The spiral structure and associated dynamics seem more complex, probably due to the accumulation of effects of the multiple (and progressively stronger) impacts and the formation of the bar and rings, as already seen in \f\ref{f_m3d}.

\ff\ref{f_LZVR3} compares the azimuthal variation of the radial velocity wave in angular momentum  at different times of the L18 model. Columns are the four different time spans specified at the beginning of the section. Overall the maximum velocities are larger for the last pericentre, as expected.
The first pericentre is dominated by two approximately flat lines of alternate $V_R$ initially\footnote{We see again some short wavelength pattern that remains from the relaxation of the initial conditions as explained in footnote \ref{f_1}.} that progressively become stepper with time and for smaller $L_Z$, very similar to Models 1. The evolution after the second and third pericentres (second and third columns) globally follows a similar behaviour with hints of the superposition of the previous wave which create substructure inside each of the negative and positive regions. At larger angular momentum we see structure not following the diagonal patterns. This could be due to larger effects in these outer disc parts due to the new encounter (not fulfilling exactly the distant tide impulsive approximation). In addition or alternatively, it could be because the signatures of the first pericentre are less phase mixed due to the longer orbital timescales, in practise superposing waves of similar wavelengths. In the last time range, the kinematics at small angular momentum seems affected by the bar, as already mentioned.  At larger angular momentum, there are more complex colour patterns than the inner diagonal stripes seen at other times and $L_Z$.

\begin{figure}
   \centering
   \includegraphics[width=0.4\textwidth]{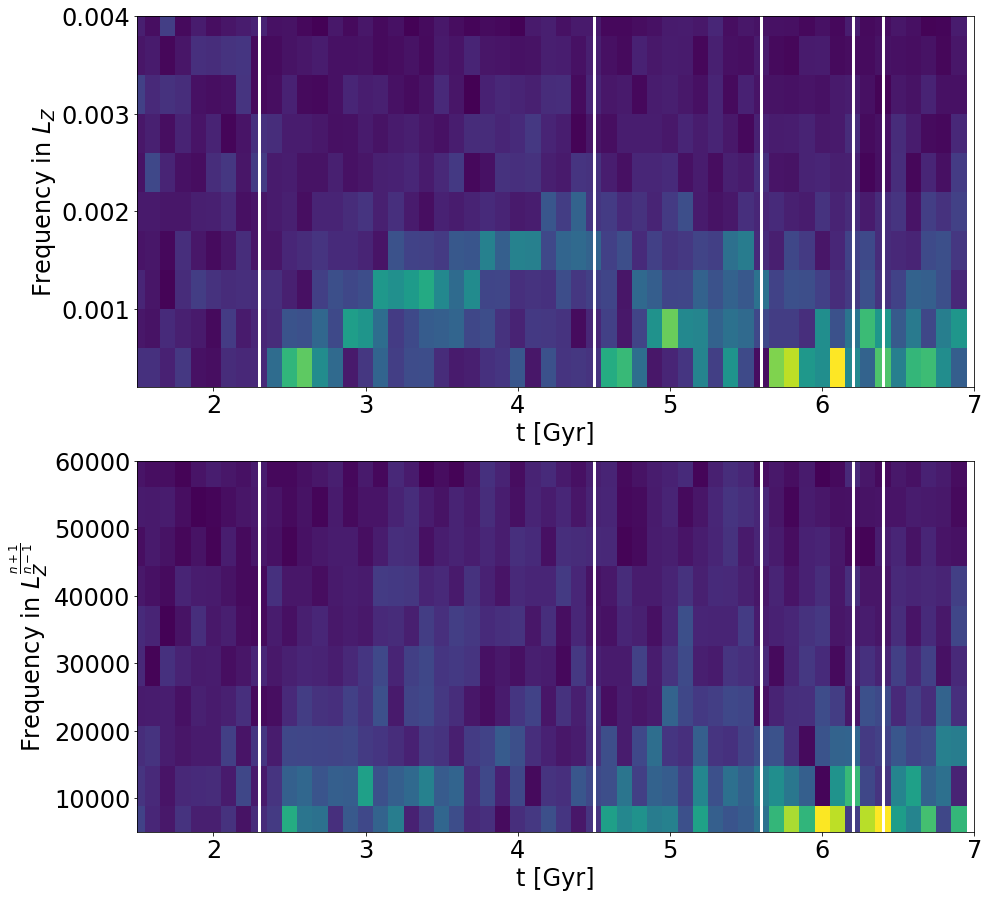}
   \caption{Frequency analysis of the $V_R$ wave for the L18 model. The plot is done as in \f\ref{f_fmodels} but we show the FT in the $L_Z$ (top) and  $L_Z^{\frac{n-1}{n+1}}$ with $n=-0.1$ (bottom).}
\label{f_fL18}%
    \end{figure}
   
Now we apply the FT to the signal in this simulation pursuing the measurement of the increase in frequency with time after the different perturbations. We arbitrarily focus on the region around $\phi=\pi$ as in previous examples. \ff\ref{f_fL18} shows the amplitude of the FT for all times. In the top panel we show the FT done in $L_Z$ space and on the bottom we use $L_Z^{\frac{n-1}{n+1}}$, motivated by the discussion of \s\ref{s_fre} and the analytical formula \e\ref{e_invL}. We note that this approach assumes that the potential is of the power-law type with a well defined $n$ (constant with $R$), while this is not exactly the case for the model L18 (see for example \citealt{GrionFilho2021} for the circular velocity curve and frequencies). This is, however, a first exploration of the potential of this methodology and not a far off approximation to realistic Galactic potentials despite its simplicity. One could also do it by using the exact frequencies of the model, which is something we can try in the future. Here we explored different values of $n$ between -0.1 and 0.1 and found that, globally, $n\sim-0.1$  retrieve higher FT amplitudes. At certain times, in particular towards the end of the simulation, $n\sim0.1$ also returns large amplitudes. This could be due to a change of the orbital frequencies at later times. Here for simplicity we show as $n=-0.1$ (bottom panel in \f\ref{f_fL18}). 

In \f\ref{f_fL18} an increase in signal can be seen coinciding with the pericentres (marked with white vertical lines) in both panels, but the signal seems clearer when the FT is done in ${L_Z}$. This could be a sign that we are not using the proper transformation to $L_Z$ as explained above. This makes the process of age dating the perturbations from the analytical formula difficult. The poor signal might also be due to the resolution of the model. In any case, it is still possible to see diagonal bands of frequency increasing linearly with time and starting at the first two pericentres. The FT amplitude also increases at times close to the last pericentres but we do not distinguish individual bands probably because of our time resolution and because the pericentres occur now very close in time and in the regime where bar effects start appearing. 

An interesting question is whether we can detect multiple perturbations with different onset times simultaneously. While some signal could be seen directly in \f\ref{f_r3a}-\ref{f_r3d} and \ref{f_LZVR3} (in some cases well after successive impacts), this is not so clear in \f\ref{f_fL18} where the start of a new diagonal line seems to wash out the signs of the previous one. At this point it is not clear whether this is a real effect due to the later perturbations being stronger, or it is a detection problem since we could not find an optimal transformation of $L_Z$ and/or binning problem. In the future one could explore doing the FT in 2D in the full $L_Z$-$\phi$ space, which most likely increases the signal, as pointed out in \s\ref{s_fre}.

\section{Data}\label{s_data}

\begin{figure*}
   \centering
   \includegraphics[width=0.99\textwidth]{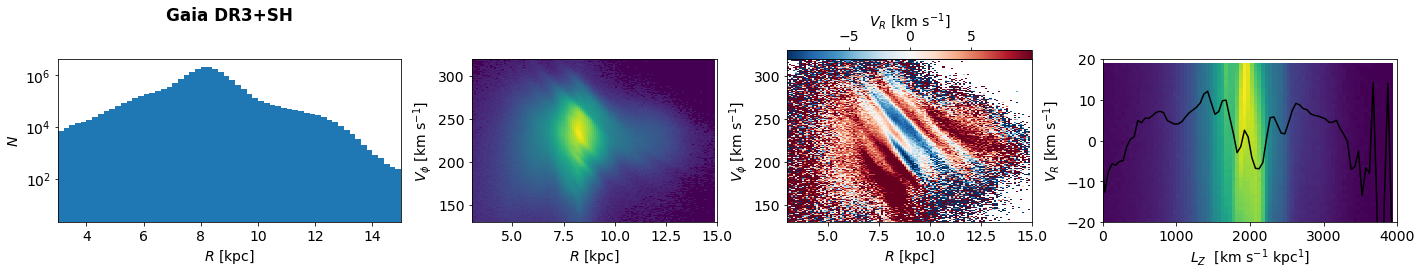}
   \caption{Ridges and waves for the MW data  ($|Z|<0.8\kpc,\ |\phi|<0.2$ rad). The different panels follow the structure of \f\ref{f_r1}.}
\label{f_ridd}%
    \end{figure*}
    \begin{figure}
   \centering
   \includegraphics[width=0.49\textwidth]{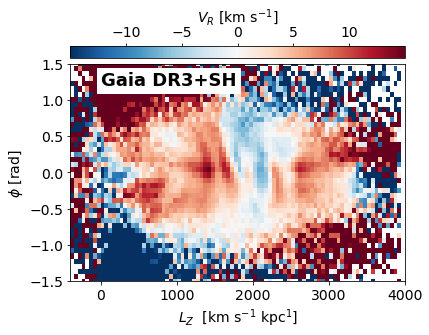}
   \caption{Azimuthal dependency of the $L_Z$-$V_R$ waves for the MW data  ($|Z|<0.8\kpc$).}
\label{f_LZVRd}%
    \end{figure}

The ridges in the $R$-$V_\phi$ projection have been studied extensively but for an easier comparison with previous models we show again some figures with real MW data using radial velocities and astrometry from DR3 \citep{Vallenari2022} and distances from StarHorse \citep{Anders2022}. We also applied the following quality cuts: i) astrometric quality selections, $\mathtt{RUWE}<1.4$ \& $\mathtt{parallax\_over\_error>5}$, and ii) selection of non-spurious solutions \citep{Rybizki2022}, $ \mathtt{fidelity\_v2}> 0.5$. We also removed stars with $\mathtt{rv\_template\_teff}\ge 8500$ K since these stars present a residual bias of a few $\kms$ in $\mathtt{radial\_velocity}$ even after some proposed corrections \citep{Blomme2022}. In addition, we applied a correction to the $\mathtt{radial\_velocity}$ to stars with $\mathtt{grvs\_mag} \ge 11$ and $\mathtt{rv\_template\_teff} <8500$ \citep{Katz2022} for which we use $  \mathtt{radial\_velocity} 
        -0.02755\times\mathtt{grvs\_mag}^2 +0.55863\times\mathtt{grvs\_mag} -2.81129$.

We further select stars in the disc with the cut at $|Z|<0.8\,\kpc$, ending up with a sample of $22\,829\,342$ 
 stars. To compute to cylindrical coordinates, we assume $R_0=8.277$ kpc, $U_{\odot} = 9.3$, $V_c+V_\odot=251$ and $W_\odot=8.59$ $\kms$ from \citet{Gravity2022} and \citet{Reid2020}, $Z_\odot=0.0208$ kpc from \citet{Bennett2019} and $V_\odot=12.24$ from \citet{Schonrich2010}.

 For \f\ref{f_ridd} we selected a wedge of $\phi_\odot\pm0.2$ rad around the Sun-Galactic Centre line. The density structure in \f\ref{f_ridd} (which is now in logarithmic scale) 
is the result of a complex convolution of the sample magnitude distribution and the different types of stars  (most of the dwarf stars being located close to the Sun) with the true underlying density profile of the Galactic disc. In the second and third columns we see multiple thin and close ridges, resembling  the later stages of a phase mixing process better than the initial ones. The existence of four minima in $V_R$ as a function of $L_Z$ (third panel) would indicate four wraps of spiral arms in the observed radial range if these were all due to a unique satellite impact but no other perturbation. We see a fifth minima at low angular momentum but at this range of $L_Z$ selection effects are more dominant (favouring eccentric orbits) and this position is closer to the region of the bar and could thus more likely be unrelated to the phenomena that we study here.

The ridges seen in the real data do not show such a clear sawtooth shape in \f\ref{f_ridd} as in some models presented in Sects. \ref{s_simple}--\ref{s_nbod}. However, this might depend on the particular plotting technique or sample used since some zigzagging seems present in Fig. 12 in \citet{Hunt2019} and Fig. 16 in \citet{Antoja2021}. 
We also note that the sawtooth shape is not as strong in the simple model with decreasing circular velocity curve (as perhaps in the MW) nor in the more realistic L18 model. The sawtooth shape also disappears when multiple perturbations coexist, which is very likely the case for our Galaxy. Additionally, the zigzag is present also in models that do not have tidally induced arms, such as those by \citet{Khanna2019} and \citet{Hunt2018}. Therefore, this shape can not be used to disentangle between models.

A wave in the $L_Z$-$V_R$ space was already detected in \citet{Friske2019} who measured two wavelengths of 1350 and 285 $\kms\kpc$ and attributed the larger one to the signatures of the bar. The new, larger sample used here shows a similar result (right panel in \f\ref{f_ridd} and \f\ref{f_LZVRd}) but much better defined oscillations and larger radial ($L_Z$) and azimuthal span. Following up on previous sections, 
we also compute the FT for the data. One could try to do a fit of \e\ref{e_invL} directly in the $L_Z$-$\phi$ space. However, the range of $\phi$ is small and we do not see a regular slope of the coloured bands in \f\ref{f_LZVRd}. This is already a warning of the higher complexity of the data compared to our models. Yet we do the FT to the $L_Z$-$V_R$ curve at $\phi=\phi_\odot$, and attempt to do a fit using \e\ref{e_Lsep} for a back-of-the-envelope calculation. To do this, we need to assume an $n$ (slope of the circular velocity curve). We considered different circular velocity curves of the MW in the literature, namely  from \citet{McMillan2017}, \citet{Sofue2020}, \citet{Eilers2019} and {\it MWpotential2014} from \citet{Bovy2015}, and we find the $n$ that best fits  each curve in  the range where we detect our signal (5.5-12 kpc). We obtain $n=$0, -0.03, -0.06, and -0.1, respectively. We note, however, that only the curves by \citet{Eilers2019} and {\it MWpotential2014} are well fitted by a single power law model. By contrast, \citet{McMillan2017}, for example, shows raising and decreasing trends with the limit at $R\approx 8\,\kpc$.

\begin{figure}
   \centering
   \includegraphics[width=0.4\textwidth]{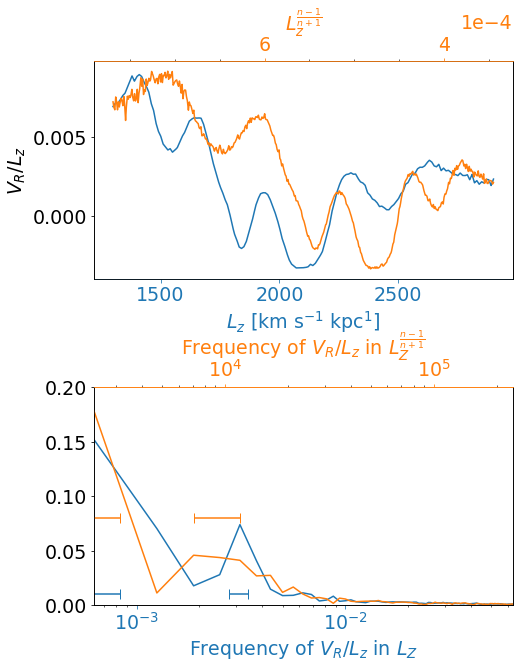}
   \caption{Frequency analysis of the $V_R$ wave in the MW data. The plot is built as in \f\ref{f_f1}. The orange curves are for $n=$0, which is one example of the different cases examined.
   The horizontal error bars indicate the uncertainty in the peaks frequency (see text for details). }
\label{f_fdata}%
    \end{figure}

In \f\ref{f_fdata} we show two curves corresponding to the $V_R$ signal (top) and FT amplitudes (bottom) in different spaces: $L_Z$ (blue, bottom axis) and $L_Z^{\frac{n-1}{n+1}}$ (orange, top axis), where we choose $n=0$ (i.e. $L_Z^{-1}$), as an example from all the above circular velocity curves. 
We find that $L_Z$ gives slightly more defined frequency peaks than $L_Z^{-1}$ and again this could be due to a bad approximation of the real potential to the power law type. In this case we also use $V_R/L_Z$ but it does not affect the results of the FT.

 In the $L_Z$ coordinate (blue) we find two defined peaks at a small and a large frequency. We note though that the peak at smaller frequency is at the limit of the first frequency point allowed by the data, and thus only an upper limit can be measured. Since our sampling of the frequencies is sparse, we only give a range of possible location of the peaks. We measured this range with 
the half-distance between the peak the consecutive frequency points. These error ranges are marked as horizontal blue error bars on the bottom panel of \f\ref{f_fdata}. The frequencies that we find correspond to wavelengths of $>1100$ and in the range of $290$-$360\,\kms\kpc$,
 thus similar to \citet{Friske2019}. 
For $L_Z^{-1}$ and in the case of the larger frequency zone, a broad flat peak is seen at approximately $10\,000$
 $(\kms\kpc)^{-1}$. 
 Again, for the smaller frequency we can only measure an upper limit.

 The limiting frequencies obtained in each case are given in \t\ref{t_fre}. In all cases, we computed the times of impact using \e\ref{e_Lsep} and assuming $V_c=239.26\,\kms$ and $R_0=8.277\,\kpc$. As expected, the obtained times are sensitive to the adopted value of $n$, with larger $n$ yielding larger times. Considering all cases, our analysis gives times of around $<0.6$ Gyr from the low frequency peak and of 0.8-2.1 Gyr from the larger frequency one.  
  We discuss these estimated times in \s\ref{s_dis2} .

\begin{table}
\setlength{\tabcolsep}{4.5pt}
\caption{Frequencies of the $V_R$ wave and times of impact from MW data. The different rows refer to when the FT is done with different coordinates ($L_Z$ and $L_Z^{\frac{n-1}{n+1}}$) as indicated in the first column. The units of the frequency are $(\kms\kpc)^{-1}$ for the first row and $((\kms\kpc)^{\frac{n-1}{n+1}})^{-1}$ for the rest. For the later cases we give also the derived impact times.    }             
\label{t_fre}      
\centering                          
\begin{tabular}{l l |r r| r r}        
\hline\hline                 
       & &  \multicolumn{2}{c|}{Low frequency}&   \multicolumn{2}{c}{High frequency} \\    \hline 
       &\multicolumn{1}{c|}{$n$} &  \multicolumn{1}{c}{Frequency}   &\multicolumn{1}{c|}{Time} &  \multicolumn{1}{c}{Frequency}   &\multicolumn{1}{c}{Time} \\   
       & &   & \multicolumn{1}{c|}{$[\Gyr]$}&   & \multicolumn{1}{c}{$[\Gyr]$}\\   
\hline                        
$L_Z$                  &          &  $<0.0009$  &    &$0.0028$ - $0.0034$ &\\  \hline 
$L_Z^{\frac{n-1}{n+1}}$&0.  &$<3500$&$<0.6$&$7100$ - $11800$&$1.3$ - $2.1$\\   
                       &-0.03&$<5300$&$<0.6$&$10600$ - $17700$&$1.2$ - $2.0$\\
                       &-0.06&$<8200$&$<0.5$&$16400$ - $27300$&$1.1$ - $1.8$\\
                       &-0.1&$<15300$&$<0.5$&$25500$ - $56100$&$0.8$ - $1.7$\\
\hline                                   
\end{tabular}
\end{table}

\section{Discussion}\label{s_dis}

\subsection{Spiral arms, ridges, waves, and frequencies}\label{s_dis1}

The response associated to the density wave theory in the tight-winding approximation (TWA) predicts that density and average radial velocity are anti-correlated inside corotation radius (i.e. the spiral arms coincide with negative radial velocities) but are positively correlated outside corotation \citep{Lin1969,Mayor1970}. The TWA arms also coincide with regions of null azimuthal velocity with respect to the average at that radius (as expected from the relation between radial and azimuthal velocities in disc orbits). 
 In \citet{Antoja2016} (see also \citealt{Siebert2012} and \citealt{RocaFabrega2014}) we found that this was indeed approximately the case in test particle simulations with potentials including TWA arms with a constant pattern speed. However, the self-consistent isolated N-body models examined in that study with strong spiral arms and with transient arms showed more complexity.
 More recently, in \citet{Eilers2020} the kinematic perturbation associated to logarithmic spiral arms was calculated through approximating orbits in a potential including the spiral perturbation. The resulting peaks of highest density, i.e. the spiral arms, coincided with locations  of zero average radial velocities. More examples of the analysis of the spiral arms' kinematics can be seen in \citet{Grand2015} and \citet{Pettitt2020} who also focused on the density wave theory, transient spirals, and spirals coexisting with bars.

The spiral structure of our models induced by an external perturber corresponds to phase-space regions with negative average radial velocities and zero azimuthal velocity with respect to the average. This is consistent with the theoretical considerations of \citet{Kalnajs1973} about kinematic waves and was observed in early N-body modelling of a galaxy perturbed by a point mass \citep{Sundelius1987} and later, for example, in \citet{Oh2008}. Here we put this in the context of spiral arms that could be tidally induced by the successive encounters of the MW and Sagittarius. This clear kinematic signature can 
potentially help to distinguish different types of spiral arms described above.

Moreover, the tidally induced spiral arms correspond to diagonal ridges with negative radial velocity $V_R$ in the $R$-$V_\phi$ projection, which does not necessarily correspond to regions of higher stellar density in this projection. Our spiral arms are also linked to the $L_Z$-$V_R$ wave, corresponding to its valleys (negative $V_R$). Since the spiral arms in density space are much  more difficult to detect in real data due to extinction and selection effects, our findings offer a potential method to infer the number of spiral arms and their location. This is different from \citet{Khoperskov2020}, where the spiral arms  were associated to regions of high density in the $R_{\rm g}$-$\phi$ plane (essentially regions of accumulation around certain angular momentum). Their spiral arms, thus, do not correspond to spiral arms necessarily seen in configuration space (see also \citealt{Hunt2020}), where we believe spiral arms must be defined. From our models we see that overdensities in certain projections of phase space do not correspond to overdensities in other projections, in particular  not in configuration space. 

In our models the number of ridges in the $R$-$V_\phi$ plane, the number of alternating changes of sign of $V_R$ in this projection, and the number of oscillations in the $L_Z$-$V_R$ wave (all these corresponding to spiral wraps) depend on the form of the planar potential and the time after the perturbation. For simple models with power law potentials, we give analytical descriptions of the number of wraps, positions of the ridges, etc, as a function of time $t$ (after the perturbation), the initial phase $\phi_0$, and the parameters of the circular velocity curves (slope $n$ and normalisation values $V_0$ and $R_0$). 
By measuring the frequency of the $L_Z$-$V_R$ wave 
and assuming a  reasonable model for the gravitational potential, one can obtain the time of onset of the perturbation. Ideally, one would try to fit the potential parameters and time of impact simultaneously. Here we also show that planar kinematic disturbances can contain signatures of multiple events as superimposed waves of different wavelengths. Our Fourier analysis allows the approximate detection of different impact times for these simple models, while there seems to be a certain detection limit for the  analysed N-body simulation,  in which there is added complexity due to other perturbations, a potential deviating from simple power laws, and low number of particles.

We note that there is a clear analogy between this and the estimation of the impact time using the phase spiral in \citet{Antoja2018} and also in using this phase-space structure to constrain the potential \citep{Widmark2021}, but our observable in this case relates to the planar potential and not the vertical one. As discussed in \citet{Laporte2019} and \citet{Hunt2021} for the phase spiral, multiple perturbations from the planar velocities could potentially be easier to identify in the outer parts of the Galaxy due to longer phase mixing times (\f\ref{f_wra}). 
 Developing a framework in which both the planar and vertical dynamics are modelled together is necessary and could provide a more robust fit to both disturbances at the same time. We now proceed to discussing the applicability of our idea to currently available Milky-Way data. 

\subsection{The case of the Milky Way}\label{s_dis2}

Our simple models with orbital integrations provide a basis to understand the dynamics of tidally induced arms and their manifestations in different phase-space projections. They have proven to be very helpful to understand what happens in more complex models such as that of L18, and potentially real disc galaxies such as the MW. One of the simplifications of our simple models is the condition of distant and impulsive impacts. A natural extension of our work could be to relax the distant-impact condition and explore different impact configurations in our toy models that could resemble the later interactions with a Sagittarius-like system, i.e. when the perturber may even cross the disc, and excite more complex structures such as asymmetric rings and spirals \citep{Appleton1996}. 
The other fundamental assumption in our simple modelling is the neglection of the effects of self-gravity of the disc and other perturbations,  mostly the bar (and possibly other types of spiral arms). More investigation is needed in order to see how these additional aspects can hamper our ability to constrain the time of the perturbations and potential of the MW. 
For example, \citet{Darling2019} show that phase-mixing timescales are longer when one considers self-gravity. Similarly, \citet{Pettitt2018} and \citet{Oh2015} show how the phase-mixing rhythms of tidally induced arms are altered by self-gravity or by the bar. 

As both the existence of the bar and of the external perturber Sagittarius are well established, the Galactic disc must be suffering from sequels
of both, the key questions being whether one of the two is the dominant mechanism,  whether the two have superposing effects - and are thus detectable as separate waves - or whether they interact with each other. \citet{Hunt2019} studied the combined effects of a bar and spiral arms with test particles and backward integration of orbits, but they considered only quasi-stationary density wave arms and transient corotating arms. In a future study, our simple modelling could be used to examine the combination of bar and tidally induced spiral arms, complemented  by a more detailed study of the later stages of the model L18 or other N-body models. In particular, one could explore how 
the FT analysis presented here could help to distinguish waves/signatures from  different dynamical processes, such as the bar. 

In the N-body simulation that we use here, the effects of the Galactic bar start to become important during the last pericentres, complicating the interpretation of the density and velocity structures due to the interaction. 
 The kinematics associated to a barred potential has been well studied in the literature. For a recent example, we refer to Fig. D.1 of \citet{Bernet2022}, which shows the $R$-$V_\phi$ projection coloured by $V_R$ for test particle simulations with different bar pattern speeds at 30 deg with respect to the bar's long axis. We see that there are clear differences between the bar and the external perturbation model. As expected, there is a limited number of ridges in the bar models: the most clear ones are those associated to corotation, Outer Lindblad Resonance (OLR) and 1:1 resonance\footnote{Other resonances, such as the 6:1, 4:1, and 3:1 resonances, are noticeable in other models \citep{Monari2019}.}. These ridges do not have  the conspicuous sawtooth shape seen in our models, and some of them have changing slopes or breaks with $R$, and their $V_R$ signal is different depending on the resonance. 
  This translates into a wave in the $L_Z$-$V_R$ space that does not have the symmetries that we see in the waves of external perturbations in the simple models. 

For the bar case, the $V_R$ wave in the $L_Z$-$\phi$ plot (similar to \fs\ref{f_LZ} and \ref{f_fL18}) can be seen in Fig. A1 of \citet{Trick2021} and  Fig. 12 of \citet{Chiba2021b}. There we see that the bar mostly creates a quadrupolar structure in $V_R$ that changes sign at the main resonances \citep[see also][]{Muhlbauer2003}. \citet{Trick2021} show that for long integration times, this structure has phase angle $\phi$ constant with $L_Z$, thus different from the external perturbation case. Only for short integration times does one see certain slopes, but this could also depend on the resonance,
 thus different from the global pattern in our models. 
These aspects should be key to consider in the comparison with the Milky-Way data (see below). However, \citet{Chiba2021b} show that this quadrupolar pattern turns into multiple bands and slopes when the bar slows down, which is a natural process in galactic dynamics. In this case, the simulations resemble more the data although additional effects might still be at play.

Finally, we note that simple barred models predict density gaps and bumps around some resonances such as the OLR (for example Fig. 9 in \citealt{Trick2021}, or Figs. 6 and 7 in \citealt{Melnik2021}). While it is well described that the different bar resonances  can create  structures in density \citep{Buta1996}, this has been hardly considered for the MW. These structures have mostly ring shape but we note that, locally, they might be confused with spiral arms. It should be explored further whether their kinematics makes them distinguishable from spiral arms.

 In any case, awaiting a clearer picture of how different dynamical effects (bar, external perturbation, etc) superpose, our simple modelling indicates two times of impact of $<0.6$ and $0.8$-$2.1$ Gyr, respectively for the small and large frequencies detected in the data. These frequencies could in principle also be related to the effects of a static bar \citep{Muhlbauer2003,Chiba2021b,Trick2021} in the small frequency case, as discussed in \cite{Friske2019}, and to a slowing bar according to \citet{Chiba2021b} in the large one. From our simple modelling, the small-frequency signal (related to the short time) could alternatively be linked to a recent passage of Sagittarius, although this is expected to happen a bit earlier in time and to have conditions far from the distant-impact hypothesis. The large-frequency peak yields times compatible with the perturbation from the previous-to-last (or earlier) pericentre \citep[e.g.][]{delaVega2015,Laporte2018,Law2010,Purcell2011}. Signatures of previous pericentres would not be washed away if Sagittarius significantly lost mass between pericentres.
 
\citet{Minchev2009} obtained a time of perturbation of 1.9 Gyr from the visual comparison of the separation between arches in the $V_R$-$V_\phi$ plane in the data and in their toy models. Their method is considerably different from ours, since we start with initial conditions that specifically mimic the ones following an impulsive and distant impact with a satellite. Furthermore, we derive an analytic formula for the oscillations in the (transformed) $L_Z$-$\phi$ space, consider the azimuthal dimension of the phase-mixing, and relate the perturbation with the tidally induced spiral arms. In any case, one of our derived perturbation times is consistent with the one obtained by \citet{Minchev2009}. 

Our derived lower limit of the impact time related to the large-frequency signal ($0.8$ Gyr) is also comparable to the times inferred from the phase spiral in \citet{Antoja2018} (500-900 Myr) in the pure phase-mixing case (although it was estimated to be 400 Myr in \citealt{Binney2018}), but this is expected to be larger in the self-gravitating case \citep{Darling2019}. The times are also compatible with direct age-dating of the phase spiral \citep{Tian2018,Laporte2019,BlandHawthorn2019}, which seems to be present in young samples (0.5-1 Gyr).

We note that the amplitude of the $V_R$ wave in the last pericentres of the L18 model is too large compared to the data ($\gtrsim 20$ vs 10 $\kms$). 
This could simply be due to the lack of exploration of different setups for Sgr's halo and disc structure in the current models. 
 It could also be connected to the existent discussion around the mass of Sagittarius at each pericentre and whether these  impacts are enough to create the phase spiral \citep[see discussion in][]{GarciaConde2022}.  This should  also be addressed in future investigations.

\section{Summary and conclusion}\label{s_con}

We have presented a series of models of tidally induced spiral arms and shown how these relate to the ridges and waves seen in certain projections of phase space, similar to those observed in the {\it Gaia} data for the MW. Our main results and conclusions are:

   \begin{enumerate}
      \item An impulsive distant tidal impact generates an initial quadrupole in velocities that leads to a two-armed spiral structure that winds up with time at an approximate rate of $\omega-1/2\kappa$. The arms progressively have smaller pitch angles and there is an increasing number of spiral wraps in a particular azimuth in the disc with time (Sect.~\ref{s_int}).
      
      \item The location of the maximum density (locus of the arms) coincides with regions where the average galactocentric radial velocity is negative, and the azimuthal velocity with respect to the average at each radius is zero (Sect.~\ref{s_int}). This is because the arms coincide with orbital phases in-between apocentre and pericentre. 
      
      \item The winding up of the spiral arms and the associated velocity pattern produce a global wave  in $V_R$ that depends on azimuth, manifesting as inclined stripes in the $L_Z$-$\phi$ plane, and as oscillations of $V_R$ as a function of $L_Z$ in a given region of the disc (Sect.~\ref{s_rid}), as observed also in the \Gaia data. The wavelength of this oscillation decreases with time (and with $L_Z$ at a fixed time). This wave is seen as a pattern of stripes with alternating signs of $V_R$ in the $R$-$V_\phi$ projection. Each stripe with negative $V_R$ corresponds to a spiral arm wrap.  
      
      \item In the $R$-$V_\phi$ plane, a sawtooth shape appears, with increasing number of oscillations with time (Sect.~\ref{s_rid}). Each oscillation is associated to a spiral wrap, but the stars that belong to the arms do not lie on the down-going ridges but on the regions with negative radial velocity in this projection. The sawtooth shape is diluted in more complex models such as N-body and when multiple mechanisms are at play (Sect.~\ref{s_nbod}).

     \item The relation between density and average velocities (and associated ridges and waves in phase space) is different for other models of spiral arms, offering a way to distinguish between different dynamical models of arms (Sect.~\ref{s_dis1}).
     
      \item With the realistic simulation of the interaction between Sagittarius and the MW, we see that the first three pericentres of Sagittarius seem close to  the impulsive and distant tide approximation in the central and intermediate regions of the disc, and thus, the dynamics follows the items above (Sect.~.\ref{s_nbod}).

      \item The rhythm at which new spiral arms wraps appear with time, and thus new ridges in the $R$-$V_\phi$ projection and oscillations of $V_R$ with $L_Z$, depends on the orbital frequencies of the disc through the radial dependence of \ok2. The number of all these phase-space features and their separation offers a means to infer at the same time the potential (essentially the circular velocity curve parameters) and the starting time and phase of the perturbation. We give analytical formulae for simple power law models that at a theoretical level (in pure phase mixing events) successfully recover the impact times and even disentangle impacts from different successive pericentres (Sect.~\ref{s_fre}). 
      
      \item For the MW we find a superposition of two different wavelengths of $>1100$ and $290$-$360\,\kms\kpc$, similar to the values of \citet{Friske2019} who detected it for the first time (Sect.~\ref{s_data}). 
      
      \item Using typical values of the circular velocity curve of the MW, we obtain times of $<0.6$ Gyr and 0.8-2.1 Gyr, associated to these two wavelengths (Sect.~\ref{s_data}). The first one is slightly smaller than the last Sagittarius pericentre and thus the long wavelength signal could be related to other mechanisms. The range of earlier times is consistent with estimated previous and previous-to-last pericentre of this galactic satellite. These would correspond to four spiral wraps in the range of the data coverage.
      
      \item The validity of this calculation is subject to the assumptions that the impact with Sagittarius was impulsive and distant, and that there is no intervention of other aspects such as self-gravity or the bar, which, on the other hand, has been shown to reproduce several of the observed kinematic features elsewhere as well (Sect.~\ref{s_dis2}).

 \end{enumerate}
 
Here we provided various models of the dynamics of tidally induced arms and a first application to the Milky Way case. However, the complex dynamics of the MW disc certainly requires more investigation.

\begin{acknowledgements}
      We thank the anonymous referee for the insightful and helpful comments. Work produced with the support of a 2021 Leonardo Grant for Researchers and Cultural Creators, BBVA Foundation. The Foundation takes no responsibility for the opinions, statements and contents of this project, which are entirely the responsibility of its authors. We thank Prof. Dr. Curtis J. Struck for useful discussions.
      TA acknowledges the grant RYC2018-025968-I funded by MCIN/AEI/10.13039/501100011033 and by ``ESF Investing in your future''. This work was (partially) funded by the Spanish MICIN/AEI/10.13039/501100011033 and by ``ERDF A way of making Europe'' by the ``European Union'' through grant RTI2018-095076-B-C21, and the Institute of Cosmos Sciences University of Barcelona (ICCUB, Unidad de Excelencia 'Mar\'{\i}a de Maeztu') through grant CEX2019-000918-M. FA acknowledges financial support from MICINN (Spain) through the Juan de la Cierva-Incorporcion programme under contract IJC2019-04862-I. MB received funding from the University of Barcelona’s official doctoral program for the development of a R+D+i project under the PREDOCS-UB grant. CL acknowledges funding from the European Research Council (ERC) under the European Union’s Horizon 2020 research and innovation programme (grant agreement No. 852839). 
      This work has made use of data from the European Space Agency (ESA) mission
{\it Gaia} (\url{https://www.cosmos.esa.int/gaia}), processed by the {\it Gaia}
Data Processing and Analysis Consortium (DPAC,
\url{https://www.cosmos.esa.int/web/gaia/dpac/consortium}). Funding for the DPAC
has been provided by national institutions, in particular the institutions
participating in the {\it Gaia} Multilateral Agreement.

\end{acknowledgements}

%
  \bibliographystyle{aa} 
  \bibliography{mybib} 
%

\end{document}